\documentclass{LMCS}

\usepackage{amsmath}
\usepackage{amssymb}
\usepackage{amsthm}
\usepackage{pst-node}
\usepackage{cmll}
\usepackage{bussproofs}
\EnableBpAbbreviations
\usepackage{url}

\usepackage{enumerate,hyperref}

\makeatletter
\def\paragraph{\@startsection{paragraph}{4}%
  \z@\z@{-\fontdimen2\font}%
  \itshape}
\makeatother

\newcommand{\ie}{\emph{i.e.}}

\newcommand{\IExcl}{{\mathord{!}}}
\newcommand{\IInt}{{\mathord{?}}}
\newcommand{\IPar}{\mathord{\parr}}
\newcommand{\ITens}{\otimes}
\newcommand{\One}{1}
\newcommand{\BoT}{{\mathord{\perp}}}

\newcommand{\cL}{\mathcal{L}}
\newcommand{\cR}{\mathcal{R}}

\newcommand{\Par}[2]{{#1}\IPar{#2}}
\newcommand{\Tens}[2]{{#1}\ITens{#2}}
\newcommand{\Orth}[1]{{#1}^\BoT}
\newcommand{\BiOrth}[1]{{#1}^{\BoT\BoT}}
\newcommand{\Excl}[1]{\IExcl{#1}}
\newcommand{\Int}[1]{\IInt{#1}}

\newcommand{\Pair}[2]{\{{#1},{#2}\}}

\newcommand{\In}{\iota}
\newcommand{\Out}{o}
\newcommand{\Cexcl}{!}
\newcommand{\Cint}{?}
\newcommand{\IParexp}{!\parr}
\newcommand{\ITensexp}{?\otimes}
\newcommand{\Gint}{\mathord?^*}
\newcommand{\Gexcl}{\mathord!^*}
\newcommand{\Sint}{\scriptstyle{\mathord?}}
\newcommand{\Sexcl}{\scriptstyle{\mathord!}}
\newcommand{\Mark}{\vbox{\hbox{$\bullet$}\vspace{-4mm}\hbox{}}}
\newcommand{\NoLab}{\tau}
\newcommand{\Labels}{\cL_\NoLab}
\newcommand{\LabelsPr}{\cL}
\newcommand{\Nets}{\mathbf\Delta}
\newcommand{\Emptynet}{\epsilon}

\newcommand{\NetRedRel}[1]{\mathrel\leadsto_{#1}}
\newcommand{\NetRedM}{\mathrel\leadsto_{\mathrm{m}}}
\newcommand{\NetRedD}{\mathrel\leadsto_{\mathrm{d}}}
\newcommand{\NetRedCp}[1]{\mathrel\leadsto_{\mathrm{c},{#1}}}
\newcommand{\NetRedNDp}[1]{\mathrel\leadsto_{\mathrm{nd},{#1}}}
\newcommand{\NetRedS}{\mathrel\leadsto_{\mathrm{s}}}
\newcommand{\NetEqD}{\mathrel\sim_{\mathrm{d}}}
\newcommand{\NetEqS}{\mathrel\sim_{\mathrm{s}}}
\newcommand{\NetLTS}{\mathbb{D}_{\cL}}
\newcommand{\LabPair}[2]{#1\overline{#2}}
\newcommand{\GenTrans}[2]{\mathrel{\stackrel{\LabPair{#1}{#2}}\longrightarrow}}
\newcommand{\NetTrans}{\GenTrans}
\newcommand{\Nat}{{\mathbb{N}}}
\newcommand{\Fmod}[2]{{#1}\langle{#2}\rangle}
\newcommand{\Rel}[1]{\mathrel{#1}}
\newcommand{\Spansymb}{\Sigma}
\newcommand{\Relspan}[1]{#1^{\mathrm\Spansymb}}
\newcommand{\Reltrans}[1]{\mathrel{#1^*}}

\newcommand{\Figspace}[1]{\hspace{#1em}}

\newcommand{\Subst}[3]{{#1}\left[{#2}/{#3}\right]}
\newcommand{\FreeNames}[1]{\operatorname{\mathsf{fn}}(#1)}
\newcommand{\ProcEmpty}{0}
\newcommand{\ProcPar}[2]{{#1}\mid{#2}}
\newcommand{\ProcNu}[2]{\nu{#1}.{#2}}
\newcommand{\ProcIn}[3]{{#1}({#2}).{#3}}
\newcommand{\ProcOut}[3]{\overline{#1}\langle{#2}\rangle.{#3}}
\newcommand{\ProcCong}{\equiv}
\newcommand{\ProcRed}{\rightarrow}
\newcommand{\ProcTrad}[2]{[{#1}]_{#2}}
\newcommand{\SoloEmpty}{0}
\newcommand{\SoloNu}[2]{({#1})\,{#2}}
\newcommand{\SoloPar}[2]{{#1}\mid{#2}}
\newcommand{\SoloIn}[2]{{#1}\,{#2}}
\newcommand{\SoloOut}[2]{\overline{#1}\,{#2}}
\newcommand{\SoloInOut}[2]{\hat{#1}\,{#2}}
\newcommand{\SoloCong}{\equiv}
\newcommand{\SoloRed}{\rightarrow}
\newcommand{\SoloCat}[1]{C_{#1}}
\newcommand{\FusionEmpty}{0}
\newcommand{\FusionNu}[2]{({#1})\,{#2}}
\newcommand{\FusionIn}[3]{\SoloIn{#1}{#2}.{#3}}
\newcommand{\FusionOut}[3]{\SoloOut{#1}{#2}.{#3}}
\newcommand{\Many}[1]{\tilde{#1}}
\newcommand{\Subject}[1]{\textit{subj}(#1)}
\newcommand{\SoloControl}{\triangleleft}
\newcommand{\SoloControlT}{\triangleleft^+}
\newcommand{\SoloControlRT}{\triangleleft^\star}
\newcommand{\SoloOrth}{\mathrel{\bot}}

\newcommand{\DINone}{S}
\newcommand{\DINtwo}{T}
\newcommand{\DINsone}{s}
\newcommand{\DINstwo}{t}

\newcommand{\SDDINRel}{\leftrightsquigarrow}
\newcommand{\SDDINRelD}{\leftrightsquigarrow_d}
\newcommand{\SDLTS}{\mathbb{S}_{\cL}}
\newcommand{\SDLTSac}{\mathbb{S}_{\cL}^\texttt{\textnormal{ac}}}
\newcommand{\SDone}{G}
\newcommand{\SDtwo}{H}
\newcommand{\SDTrans}{\GenTrans}
\newcommand{\SDred}[2]{#1[#2]}
\newcommand{\idgraph}[1]{#1_{\text{id}}}
\newcommand{\NdSDred}[2]{\idgraph{#1}[#2]}
\newcommand{\node}[1]{\textnormal{\textsf{#1}}}

\newcommand{\Prot}[1]{\textnormal{\texttt{#1}}}
\newcommand{\ProtX}{\Prot{X}}
\newcommand{\ProtE}{\Prot{S}}
\newcommand{\ProtR}{\Prot{R}}

\def\doi{6 (3:11) 2010}
\lmcsheading%
{\doi}
{1--34}
{}
{}
{Oct.~20, 2008}
{Sep.~\phantom01, 2010}
{}

\begin{document}

\title[Acyclic Solos and Differential Interaction Nets]{Acyclic Solos and Differential Interaction Nets}
\author[T.~Ehrhard]{Thomas Ehrhard}
\address{Preuves Programmes Syst\`emes,
        CNRS -- Universit\'e Paris 7,
        France}
\email{Thomas.Ehrhard@pps.jussieu.fr, Olivier.Laurent@ens-lyon.fr}
\thanks{This work has been partially funded by the French ANR projet blanc ``Curry Howard pour la Concurrence'' CHOCO ANR-07-BLAN-0324.}
\author[O.~Laurent]{Olivier Laurent}



\begin{abstract}
  We present a restriction of the solos calculus which is stable under reduction and expressive enough to contain an encoding of the pi-calculus. As a consequence, it is shown that equalizing names that are already equal is not required by the encoding of the pi-calculus. In particular, the induced solo diagrams bear an acyclicity property that induces a faithful encoding into differential interaction nets. This gives a (new) proof that differential interaction nets are expressive enough to contain an encoding of the pi-calculus.

  All this is worked out in the case of finitary (replication free) systems without sum, match nor mismatch.
\end{abstract}

\maketitle

\section*{Introduction}

The question of extending the Curry-Howard correspondence (between the $\lambda$-calculus and intuitionistic logic) to concurrency theory is a long-standing open problem. We developed in a previous paper~\cite{pidinic} a translation between the $\pi$-calculus~\cite{pi} and differential interaction nets~\cite{din}. This has shown that differential linear logic --- a logical system whose sequent calculus~\cite{pidinic} has been obtained by the first author from a precise analysis of some denotational models of linear logic based on vector spaces~\cite{kothe} --- is a reasonable candidate for a Curry-Howard correspondence with concurrent computation, since differential interaction nets appear to be expressive enough to represent key concurrency primitives as provided by the $\pi$-calculus.

There are other works trying to relate variants of linear logic proof-nets with process algebras (such as~\cite{Abramsky93,BellinScott94,BeffaraMaurel05,FaggianMaurel05,Mazza05,picppn}) or to work on graphical syntaxes for encoding the $\pi$-calculus (see~\cite{pinets,solodiag,JensenMilner04} for example). However, even if some of them try to build bridges with (linear) logic, they require some ad hoc graphical constructions without logical foundations. Our goal is to build on differential interaction nets coming with the logical justification of differential linear logic.

Let us tell a bit more about the genesis of our translation. We discovered the notion of \emph{communication areas} as particular differential interaction nets able to represent some communication primitives. However, due to the asynchronous flavour of differential interaction nets, it was not immediate that additional constructions such as prefixing could be easily encoded. This led us to have a look at the solos calculus.

The \emph{solos calculus}~\cite{solos} has allowed to prove how action prefixes (thus sequentiality constraints) can be encoded in a calculus without prefixes (results of this kind also appear in~\cite{combinpi} and are going much further than in the asynchronous $\pi$-calculus~\cite{HondaK:ocac,BoudolG:aynatpic}). This is done by encoding the $\pi$-calculus into the solos calculus~\cite{solos}. 
From this point of view, the solos calculus can be seen as a low level concurrent language: even more basic than the $\pi$-calculus.

The main goal of the present paper is to stress the very close relation which appears between \emph{differential interaction nets}~\cite{din} (the graphical syntax for differential linear logic) and \emph{solo diagrams}~\cite{solodiag} (the graphical syntax for the solos calculus).

\bigskip

Following the ingredients of~\cite{pidinic}, it is possible to give a simple representation of solo diagrams as differential interaction nets (based on an implementation of nodes as communication areas). Such a ``static'' correspondence is not very interesting if there is no ``dynamic'' counterpart in the reduction semantics. And this is where troubles come into the picture... The reduction semantics \emph{almost} match...
There is a mismatch between the solos calculus and communication areas with respect to the identification of an occurrence of name with another occurrence of the same name during reduction (let us call this phenomenon, \emph{self-identification}).
Note that such an identification never occurs in the $\pi$-calculus. Since name passing is handled by the substitution of a bound name by \emph{another} name.
As in the \emph{fusion calculus}~\cite{fusion}, the solos calculus defines communication by unification of names, so that two occurrences of a name could be identified by reduction. If $x$ has to be identified with itself, it is just considered as a dummy operation in a unification setting since we already have $x=x$.
The situation is different with differential interaction nets since communication areas keep track of such an identification through an explicit link connecting the communication area associated with $x$ to itself. Reduction is not able to simplify such a link in order to give back a communication area. One gets some more complicated structure.

In~\cite{pidinic} we were directly working with the $\pi$-calculus (thus with name passing by substitution, thus without self-identification). This way a direct interpretation of the whole (finitary) $\pi$-calculus into differential interaction nets was possible.
The approach we propose here is different: we look for a restriction of the solos calculus (avoiding self-identification) for which the static correspondence with differential interaction nets also works at the dynamic level. This is the \emph{acyclic solos calculus}.
This requires us to restrict the source calculus but allows us to deal with a very natural translation.

\bigskip

The acyclic solos calculus is obtained from a precise analysis of the translation of the $\pi$-calculus into the solos calculus. The main idea is to be able to isolate, inside the unification mechanism which is a symmetric operation (when one unifies a name $x$ with a name $y$), a ``flow'' from one side to the other (from $x$ to $y$ for example) as it happens in a substitution mechanism (in the reduction of $\ProcPar{\ProcOut u{x}P}{\ProcIn u{y}Q}$ in the $\pi$-calculus, one can consider a flow going from $x$ to $y$, with the binding occurrence $y$ ``controlling'' all the other occurrences of $y$ in $Q$).
The first ingredient in our definition of the acyclic solos calculus is a simple typing system allowing us to label all the object occurrences of names of a solos term with two \emph{protocols} (\ProtE{} (send) and \ProtR{} (receive)) in a kind of uniform way. The main consequence of the typing system is to break symmetry in reduction/unification: when two occurrences of names have to be unified one is an $\ProtE$-occurrence and the other one is an $\ProtR$-occurrence. However this does not impose important structural or behavioural constraints on protocols. This is given by the second part of the definition of the calculus: the \emph{acyclicity conditions} (\ref{enumlinear}--\ref{enumbound}). The justification for these conditions is to mimic the structure of the input prefix $\ProcIn u{y}Q$ of the $\pi$-calculus inside the solos calculus: exactly one binding occurrence $y$ (this will be the $\ProtR$-occurrence of the name $y$ in the solos term), all the other occurrences (thus $\ProtE$-occurrences for object occurrences in acyclic solos) are coming ``sequentially after'' the binding one, ...
Informally, all this together leads to a name passing by unification which is very close to a substitution mechanism if we understand the unification of an $\ProtE$-occurrence with an $\ProtR$-occurrence as a substitution flow from the $\ProtE$-occurrence to the $\ProtR$-occurrence (and then all the other occurrences of this name in $\ProtE$-position). Formally, we prove that self-identification never occurs in the acyclic solos calculus.

We also show that the translation of the (finitary) $\pi$-calculus into the solos calculus following~\cite{solos} has its range contained in the acyclic solos calculus. This proves that our restriction of the solos calculus still has a reasonable expressive power. Moreover this gives us an alternative proof (with respect to the work presented in~\cite{pidinic}) of the existence of a translation of the finitary $\pi$-calculus into differential interaction nets.

\bigskip

The first part of the paper is devoted to the introduction of the required calculi ($\pi$-calculus and solos calculus in Section~\ref{seccalc}, solo diagrams in Section~\ref{secsolodiag}, and differential interaction nets in Section~\ref{secdin} which is almost copy-pasted from~\cite{pidinic}). In Section~\ref{sectrans} we elaborate on the material presented in~\cite{pidinic} to define the simple static translation of solo diagrams~\cite{solodiag} (the graphical syntax for the solos calculus) into differential interaction nets. We discuss the problems arising at the dynamic level, and we give a sufficient condition on solo diagrams for the translation into differential interaction nets to be a bisimulation: an occurrence of a name should never be unified with another occurrence of the same name (\ie{} no self-identification).

The main technical contribution of the paper comes in the last part (Section~\ref{secacsolos}). Since the property that there is no self-identification in the redexes of a solos term is of course \emph{not} preserved under the reduction of solos, we have to find a more clever property. We define the \emph{acyclic solos calculus} through a typing system for assigning protocols and the acyclicity conditions (\ref{enumlinear}--\ref{enumbound}).
We prove this restriction to be well behaved with respect to the reduction of the solos calculus. We show that the translation of a $\pi$-term is always an acyclic solos term, showing the expressiveness of the system. Finally we prove that the sufficient condition introduced in Section~\ref{sectrans} is fulfilled by solo diagrams corresponding to acyclic solos terms, showing that we obtain a bisimulation with respect to differential interaction nets.

\section{The \texorpdfstring{$\pi$}{pi}-calculus and the solos calculus}\label{seccalc}

In this section we recall the definition of the $\pi$-calculus and of the solos calculus we are going to use. We also recall the translation from $\pi$ to solos given in~\cite{solos}.

In the $\pi$-calculus~\cite{pi}, the two basic components of communication are the output and input prefixes $\ProcOut u{x}P$ and $\ProcIn v{y}Q$ (with occurrences of $y$ bound in $Q$). Communication can occur if $u=v$ and in this case, it is obtained by substituting $y$ by $x$ (in $Q$). Moreover these prefix constructions induce a sequential behaviour: the continuation $P$ (resp.\ $Q$) cannot interact with other agents before the output on $u$ (resp.\ the input on $v$) has been triggered.

The fusion calculus~\cite{fusion} provides a generalization with communication modelized by unification. This makes sending and receiving perfectly symmetric and thus the distinction between $\FusionOut u{x}P$ and $\FusionIn u{x}P$ is purely formal. Interaction is just constrained to occur between two dual entities $u$ and $\overline{u}$. The symmetry is broken when one translates $\pi$-terms into fusion terms by: $\ProcIn u{x}P\mapsto\FusionNu{x}{\FusionIn u{x}P}$ (and unification restrained to the image of this translation becomes substitution by properties of the binding restriction $\FusionNu{x}P$). Notice that the sequentiality aspect of prefixing remains.

Based on the unification mechanism provided by the fusion calculus, the solos calculus~\cite{solos} is free from explicit sequentiality constructions: the fusion prefixes are restricted to dummy continuations $\FusionOut u{x}\FusionEmpty$ and $\FusionIn u{x}\FusionEmpty$ (where $\FusionEmpty$ is an inactive process). It is shown in~\cite{solos} that dyadic solos $\SoloOut u{xy}$ and $\SoloIn u{xy}$ are as expressive as the whole fusion calculus.

\paragraph{Comments on chosen calculi.}
Since our goal is to focus on prefixing and sequentiality, we deal with calculi without replications nor recursive definitions, without match/mismatch and without sums (see Conclusion for additional comments).

We do not want to spend time to deal with arbitrary arities in the calculi we consider.
This is why we only consider \emph{monadic} $\pi$-terms. There are three reasons for that: it makes the presentation simpler, it does not lead to a loss of expressiveness, and finally the polyadic case has already been considered in~\cite{pidinic}.
As a consequence (see the translation in Section~\ref{sectranspisolos}), we are led to consider a \emph{triadic} solos calculus (all the names are of arity exactly $3$) and \emph{triadic} solo diagrams (all the multiedges are of arity exactly $3$).
The more general case of arbitrary arities could easily be obtained by introducing the appropriate sortings on the various calculi.

\subsection{The \texorpdfstring{$\pi$}{pi}-calculus}

The terms of the (monadic, finitary) $\pi$-calculus are given by:
\begin{equation*}
\begin{array}{ccccccccccc}
  P & ::= & \ProcEmpty &\mid& \ProcIn u{x}P &\mid& \ProcOut u{x}P &\mid& (\ProcPar{P}{P}) &\mid& \ProcNu x{P}
\end{array}
\end{equation*}
where both $\ProcIn u{x}P$ and $\ProcNu x{P}$ bind occurrences of $x$ in $P$.

The \emph{structural congruence} on $\pi$-terms is the least congruence containing $\alpha$-equivalence and:
\begin{align*}
  \ProcPar\ProcEmpty P &\ProcCong P\\
  \ProcPar PQ &\ProcCong \ProcPar QP\\
  \ProcPar{(\ProcPar PQ)}R &\ProcCong \ProcPar P{(\ProcPar QR)}\\
  \ProcNu x{\ProcNu y P} &\ProcCong \ProcNu y{\ProcNu x P}\\
  \ProcNu x\ProcEmpty &\ProcCong \ProcEmpty\\
  \ProcPar{(\ProcNu xP)}Q &\ProcCong \ProcNu x{(\ProcPar PQ)} &\text{if $x\notin\FreeNames Q$}
\end{align*}

The \emph{reduction semantics} of the $\pi$-calculus is given by:
\begin{gather*}
  \AXC{}
  \UIC{$\ProcPar{\ProcOut u{x}P}{\ProcIn u{y}Q} \ProcRed \ProcPar P{\Subst Q{x}{y}}$}
  \DP \\[2ex]
  \AXC{$P\ProcRed Q$}
  \UIC{$\ProcPar PR\ProcRed\ProcPar QR$}
  \DP
\qquad\qquad
  \AXC{$P\ProcRed Q$}
  \UIC{$\ProcNu xP\ProcRed\ProcNu xQ$}
  \DP
\qquad\qquad
  \AXC{$P\ProcCong P'$}
  \AXC{$P'\ProcRed Q'$}
  \AXC{$Q'\ProcCong Q$}
  \TIC{$P\ProcRed Q$}
  \DP
\end{gather*}

\subsection{The solos calculus}\label{secsolos}

Introduced in~\cite{solos}, the goal of the solos calculus is to prove the expressiveness of a calculus without prefix construction.

The terms of the (triadic) solos calculus are given by:
\begin{equation*}
\begin{array}{ccccccccccc}
  P & ::= & \SoloEmpty &\mid& \SoloIn u{x_1x_2x_3} &\mid& \SoloOut u{x_1x_2x_3} &\mid& (\SoloPar{P}{P}) &\mid& \SoloNu x{P}
\end{array}
\end{equation*}
where $\SoloNu x{P}$ binds occurrences of $x$ in $P$.

The \emph{structural congruence} is the least congruence containing $\alpha$-equivalence and:
\begin{align*}
  \SoloPar\SoloEmpty P &\SoloCong P\\
  \SoloPar PQ &\SoloCong \SoloPar QP\\
  \SoloPar{(\SoloPar PQ)}R &\SoloCong \SoloPar P{(\SoloPar QR)}\\
  \SoloNu x{\SoloNu y P} &\SoloCong \SoloNu y{\SoloNu x P}\\
  \SoloNu x\SoloEmpty &\SoloCong \SoloEmpty\\
  \SoloPar{(\SoloNu xP)}Q &\SoloCong \SoloNu x{(\SoloPar PQ)} &\text{if $x\notin\FreeNames Q$}
\end{align*}
This equivalence allows us to present terms in the solos calculus in canonical forms: either the $\SoloEmpty$ process or a bunch of scope constructions $\SoloNu {x_1}{\SoloNu {x_2}{\dots}}$ followed by solos in parallel.

The \emph{reduction semantics} of the solos calculus is given by ($\Many z$ stands for $z_1\dots z_n$):
\begin{equation*}
  \AXC{}
  \UIC{$\SoloNu {\Many z}{(\SoloPar{\SoloPar{\SoloOut u{x_1x_2x_3}}{\SoloIn u{y_1y_2y_3}}}P)} \SoloRed P\sigma$}
  \DP
\end{equation*}
\begin{center}
\begin{minipage}{10cm}
where $\sigma$ is a most general unifier of $x_1x_2x_3$ and $y_1y_2y_3$, such that exactly the names $w$ in $\Many z$ are modified, \ie{} satisfy $\sigma(w)\neq w$ (in particular, in each equivalence class of names induced by unification, at most one name is free).
\end{minipage}
\end{center}
\begin{equation*}
  \AXC{$P\SoloRed Q$}
  \UIC{$\SoloPar PR\SoloRed\SoloPar QR$}
  \DP
\qquad\qquad
  \AXC{$P\SoloRed Q$}
  \UIC{$\SoloNu xP\SoloRed\SoloNu xQ$}
  \DP
\qquad\qquad
  \AXC{$P\SoloCong P'$}
  \AXC{$P'\SoloRed Q'$}
  \AXC{$Q'\SoloCong Q$}
  \TIC{$P\SoloRed Q$}
  \DP
\end{equation*}
An alternative (but equivalent) definition of this reduction semantics is given in~\cite{solos} together with additional explanations.

For example, in $\SoloNu x{\SoloNu y{\SoloNu z{\SoloNu w{(\SoloPar{\SoloOut u{uxy}}{\SoloPar{\SoloIn u{zww}}{\SoloOut v{zuy}}})}}}}$, the only possible reduction is between $\SoloOut u{uxy}$ and $\SoloIn u{zww}$. It induces the identifications $u=z$, $x=w$ and $y=w$, thus two equivalence classes $\{u,z\}$ and $\{x,y,w\}$. In the first one, $u$ is free and thus the only possibility is to map $z$ to $u$. In the second one, all the elements are bound, we choose one of them: $y$ for example (the other choices would lead to structurally congruent results). We consider the unifier containing $z\mapsto u$, $x\mapsto y$, $w\mapsto y$ and which is the identity on the other names.
We obtain the reduct $\SoloNu y{\SoloOut v{uuy}}$.

\subsection{From \texorpdfstring{$\pi$}{pi}-terms to solos}\label{sectranspisolos}

In~\cite{solos}, the authors give different translations of the fusion calculus~\cite{fusion} into solos. We are going to focus on one of them (the one which does not introduce matching). By pre-composing this translation with the canonical embedding of the $\pi$-calculus into the fusion calculus: $\ProcIn u{x}P\mapsto\FusionNu{x}{\FusionIn u{x}P}$, we obtain the translation of the $\pi$-calculus into solos that we present here.

A $\pi$-term $P$ is translated as $\ProcTrad{P}{}$:
\begin{align*}
  \SoloCat v &:= \SoloNu z {\SoloIn v {zzv}} \\[2ex]
  \ProcTrad{\ProcEmpty}{v} &:= \SoloEmpty \\
  \ProcTrad{\ProcIn u{x}P}{v} &:= \SoloNu w{\SoloNu y{(\SoloPar{\SoloOut v{uwy}}{\SoloPar{\SoloCat y}{\SoloNu{x}{\SoloNu{v'}{(\SoloPar{\SoloIn w{xvv'}}{\ProcTrad{P}{v'}})}}}})}} \\
  \ProcTrad{\ProcOut u{x}P}{v} &:= \SoloNu w{\SoloNu y{(\SoloPar{\SoloOut v{uwy}}{\SoloPar{\SoloCat y}{\SoloNu{v'}{(\SoloPar{\SoloOut w{xv'v}}{\ProcTrad{P}{v'}})}}})}} \\
  \ProcTrad{\ProcPar{P}{Q}}{v} &:= \SoloPar{\ProcTrad{P}{v}}{\ProcTrad{Q}{v}} \\
  \ProcTrad{\ProcNu x{P}}{v} &:= \SoloNu x{\ProcTrad{P}{v}} \\[2ex]
  \ProcTrad{P}{} &:= \SoloNu v{(\SoloPar{\ProcTrad{P}{v}}{\SoloCat v})}
\end{align*}
As shown in~\cite{solos}, this encoding is adequate with respect to weak barbed congruence $\approx$: $\ProcTrad{P}{}\approx\ProcTrad{Q}{}$ implies $P\approx Q$.

The $\pi$-term $\ProcPar{\ProcNu x{(\ProcPar{\ProcOut ux\ProcEmpty}{\ProcIn xy\ProcEmpty})}}{\ProcIn uz{\ProcOut zt\ProcEmpty}}$ is translated as a solos term which is structurally congruent to $\SoloNu v{(\SoloPar{\SoloPar{\SoloNu x{(\SoloPar{P}{Q})}}{R}}{\SoloCat v})}$ with:
\begin{align*}
P&=\SoloNu w{\SoloNu y{(\SoloPar{\SoloOut v{uwy}}{\SoloPar{\SoloCat y}{\SoloNu{v'}{\SoloOut w{xv'v}}}})}} \\
Q&=\SoloNu w{\SoloNu {y'}{(\SoloPar{\SoloOut v{xwy'}}{\SoloPar{\SoloCat {y'}}{\SoloNu{y}{\SoloNu{v'}{\SoloIn w{yvv'}}}}})}} \\
R&=\SoloNu w{\SoloNu y{(\SoloPar{\SoloOut v{uwy}}{\SoloPar{\SoloCat y}{\SoloNu{z}{\SoloNu{v'}{(\SoloPar{\SoloIn w{zvv'}}{\SoloNu {w}{\SoloNu y{(\SoloPar{\SoloOut {v'}{zwy}}{\SoloPar{\SoloCat y}{\SoloNu{v''}{\SoloOut {w}{tv''v'}}}})}}})}}}})}}
\end{align*}
By applying reduction, we obtain in particular the following reducts (up to structural congruence):
\begin{gather*}
  \SoloNu v{(\SoloPar{\SoloNu x{(\SoloPar{(\SoloPar{\SoloCat v}{\SoloNu{v'}{\SoloOut u{xv'v}}})}{Q})}}{R})} \\
  \SoloNu v{(\SoloPar{\SoloNu x{(\SoloPar{(\SoloNu{v'}{\SoloOut u{xv'v}})}{Q})}}{\SoloPar{\SoloCat v}{\SoloNu{z}{\SoloNu{v'}{(\SoloPar{\SoloIn u{zvv'}}{\SoloNu {w}{\SoloNu y{(\SoloPar{\SoloOut {v'}{zwy}}{\SoloPar{\SoloCat y}{\SoloNu{v''}{\SoloOut {w}{tv''v'}}}})}}})}}}})} \\
  \SoloNu v{\SoloNu x{(\SoloPar{Q}{\SoloPar{\SoloCat v}{(\SoloNu {w}{\SoloNu y{(\SoloPar{\SoloOut {v}{xwy}}{\SoloPar{\SoloCat y}{\SoloNu{v''}{\SoloOut {w}{tv''v}}}})}})}})}} \\
  \SoloNu v{\SoloNu x{(\SoloPar{\SoloPar{\SoloCat {v}}{\SoloNu{y}{\SoloNu{v'}{\SoloIn x{yvv'}}}}}{(\SoloNu {w}{\SoloNu y{(\SoloPar{\SoloOut {v}{xwy}}{\SoloPar{\SoloCat y}{\SoloNu{v''}{\SoloOut {w}{tv''v}}}})}})})}} \\
  \SoloNu v{\SoloNu x{(\SoloPar{\SoloNu{y}{\SoloNu{v'}{\SoloIn x{yvv'}}}}{\SoloPar{\SoloCat v}{\SoloNu{v''}{\SoloOut {x}{tv''v}}}})}} \\
  \SoloNu v{\SoloCat v}
\end{gather*}

\section{Solo diagrams}\label{secsolodiag}

To make the relation with differential interaction nets simpler, we go through the graphical syntax associated with the solos calculus: solo diagrams~\cite{solodiag}.

A (triadic) \emph{solo diagram} is given by a finite set of nodes and a finite multiset of (ternary) multiedges (directed edges with a list of three nodes as source and one node as target). A node is tagged as either \emph{free} or \emph{bound}.
A multiedge is tagged as either \emph{input} or \emph{output}. Any node must be a source or a target of a multiedge.

\subsection{Reduction}

The reduction of solo diagrams is given:
\begin{enumerate}[$\bullet$]
\item by choosing two multiedges $e_1$ and $e_2$ of opposite polarities (one input and one output) with the same target (we call them \emph{dual multiedges}) and with respective sources $[\node{n}_1,\node{n}_2,\node{n}_3]$ and $[\node{m}_1,\node{m}_2,\node{m}_3]$,
\item and by identifying the two nodes $\node{n}_1$ and $\node{m}_1$, the two nodes $\node{n}_2$ and $\node{m}_2$, and the two nodes $\node{n}_3$ and $\node{m}_3$.
The reduction is allowed to occur only if it does not identify two free nodes. Moreover when a free node is identified with a bound node, the obtained node is free and when two bound nodes are identified, the obtained node is bound.
\end{enumerate}
The chosen multiedges are erased and nodes which are not anymore source or target of a multiedge are also removed. This can be applied to both free and bound nodes.

During a reduction step, the number of multiedges decreases by two and the number of nodes decreases or remains the same.

\begin{figure}
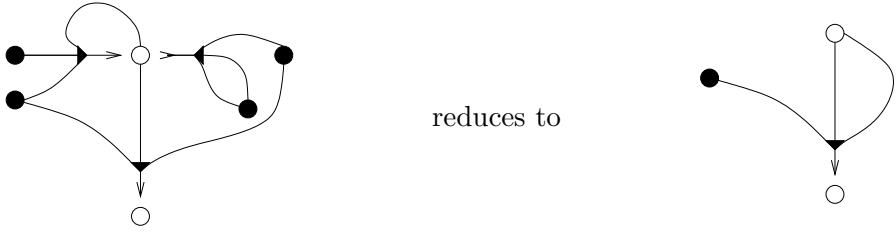

  \centering
\begin{tabular}[c]{c}
\scalebox{0.75}{\input{soloex1.pstex_t}}
\end{tabular}
\qquad\qquad reduces to \qquad\qquad
\begin{tabular}[c]{c}
\scalebox{0.75}{\input{soloex2.pstex_t}}
\end{tabular}
  \caption{A reduction in solo diagrams.}\label{figexsdiag}
\end{figure}

In the graphical representation, we draw free nodes as white dots, bound nodes as black dots, output edges with an outgoing arrow and input edges with an ingoing arrow.
An example is given in Figure~\ref{figexsdiag}.

\subsection{Relation with solos}

A term of the solos calculus can easily be translated into a solo diagram. Free names are translated as free nodes, bound names as bound nodes and solos as multiedges:
\begin{enumerate}[$\bullet$]
\item The $\SoloEmpty$ term is translated as the empty graph with no edge.
\item The solo $\SoloIn u{x_1x_2x_3}$ is translated as the graph with free nodes corresponding to elements of $\{u,x_1,x_2,x_3\}$, and with one input multiedge with source $[\node{x}_1,\node{x}_2,\node{x}_3]$ and target $\node{u}$ (where $\node{x}_i$ is the node corresponding to the name $x_i$ and $\node{u}$ is the node corresponding to the name $u$).
\item The solo $\SoloOut u{x_1x_2x_3}$ is translated in the same way with an output multiedge.
\item The parallel composition $\SoloPar{P}{Q}$ is obtained by ``graph union'': union of the sets of nodes (\ie{} nodes corresponding to the same name are identified) and sum of the multisets of multiedges.
\item The restriction $\SoloNu x{P}$ corresponds to turning the free node $\node{x}$ corresponding to the name $x$ (if any) into a bound node.
\end{enumerate}
As shown in~\cite{solodiag}, two solos terms are structurally congruent if and only if the corresponding solo diagrams are isomorphic. Moreover reduction in solo diagrams reflects faithfully the reduction of the solos calculus.

The two solo diagrams of Figure~\ref{figexsdiag} correspond to
$\SoloNu x{\SoloNu y{\SoloNu z{\SoloNu w{(\SoloPar{\SoloOut
          u{uxy}}{\SoloPar{\SoloIn u{zww}}{\SoloOut v{zuy}}})}}}}$,
respectively, $\SoloNu y{\SoloOut v{uuy}}$ (the examples used in Section~\ref{secsolos}).

\subsection{Labeled transition system}\label{secsdlts}

We follow the same methodology as in~\cite{pidinic} by using labels to distinguish the actions of a process. In a term syntax as in~\cite{pidinic}, labels are put on prefixes (or solos). In the graphical syntax of solo diagrams, we put labels on multiedges (since multiedges correspond to solos).

We fix a countable set $\LabelsPr$ of labels to be used for all our labeled transition systems.

A solo diagram is \emph{labeled} if its multiedges are equipped with different labels belonging to $\LabelsPr$.

\begin{defi}[The transition system $\SDLTS$]
The objects of the \emph{labeled transition system $\SDLTS$} are labeled solo diagrams and transitions are labeled by pairs of distinct elements of $\LabelsPr$. Let $\SDone$ and $\SDtwo$ be two labeled solo diagrams, $\SDone\SDTrans lm\SDtwo$ if $\SDtwo$ is obtained from $\SDone$ by applying a reduction step to the input multiedge labeled $l$ and to the output multiedge labeled $m$ (the labels of the remaining multiedges of $\SDtwo$ must be the same as for the corresponding ones in $\SDone$).
\end{defi}

This is quite different from what is usually done for defining transition systems for a process algebra~\cite{pi}. The standard approach aims at analyzing the possible interactions of a process with its environment. Then bisimulation is used to show that two processes have the same ``external'' behaviour.
Our goal is to use bisimulation to compare the internal behaviours of solo diagrams and of differential interaction nets. We want to illustrate that differential interaction nets are sufficiently expressive for simulating concurrency and mobility. This requires the transition systems we use for bisimulation to carry precise informations about reductions of processes, in particular about the sub-entities taking part into the interactions.

\subsection{Solo diagrams with identifications}

In order to compare solo diagrams with differential interaction nets, it will be useful to decompose the reduction of solo diagrams by introducing the notion of \emph{solo diagrams with identifications}.

\begin{defi}[Solo diagram with identifications]
\emph{Solo diagrams with identifications} are solo diagrams equipped with a finite set of undirected edges (usual binary edges, not multiedges, which connects nodes of the solo diagram). These edges are called \emph{identification edges}.
\end{defi}

We can decompose the reduction of solo diagrams we have presented above by using identification edges. A reduction step for a solo diagram $\SDone$ is defined as follows:
\begin{enumerate}[(R1)]
\item choose two dual multiedges $e_1$ and $e_2$ (\ie{} of opposite polarities and with the same target) with respective sources $[\node{n}_1,\node{n}_2,\node{n}_3]$ and $[\node{m}_1,\node{m}_2,\node{m}_3]$;
\item\label{stepident} build the solo diagram with identifications $\SDred{\SDone}{e_1,e_2}$ obtained from $\SDone$ by erasing $e_1$ and $e_2$ and by introducing three identification edges: between $\node{n}_1$ and $\node{m}_1$, between $\node{n}_2$ and $\node{m}_2$, and between $\node{n}_3$ and $\node{m}_3$;
\item\label{stepcontract} contract the graph $\SDred{\SDone}{e_1,e_2}$ by (repeatedly) choosing an identification edge and by identifying the two (or one) nodes it connects if at least one of them is bound.
\end{enumerate}
The reduction succeeds (\ie{} is a valid reduction of solo diagrams) if we reach a solo diagram with no remaining identification edge. It means in particular that step~(\ref{stepident}) does not introduce identification edges between two free names.

In the graphical representation, we draw identification edges as dashed edges.

\begin{figure}
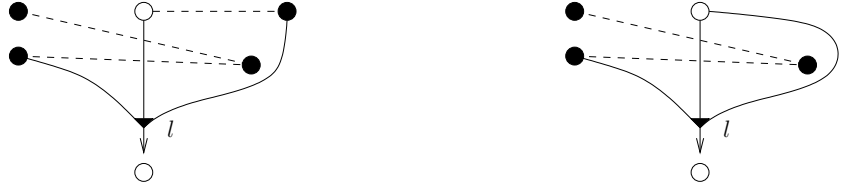

  \centering
\begin{tabular}[c]{c}
\scalebox{0.75}{\input{soloex3.pstex_t}}
\end{tabular}
\qquad\qquad\qquad\qquad
\begin{tabular}[c]{c}
\scalebox{0.75}{\input{soloex4.pstex_t}}
\end{tabular}
\caption{An (\ref{stepcontract}) reduction step in solo diagrams with identifications.}\label{figexsdid}
\end{figure}

If we refine the reduction of the example of Figure~\ref{figexsdiag} by means of solo diagrams with identifications, the results of step~(\ref{stepident}) and then of one application of step~(\ref{stepcontract}) are in Figure~\ref{figexsdid}.

\section{Differential Interaction Nets}\label{secdin}

\subsection{The general formalism of interaction nets}\label{sec:general-nets}

We recall the general (graphical) syntax of interaction nets, as introduced
in~\cite{pnin}. See also~\cite{phdmazza,din} for more details. 

\paragraph{Statics of interaction nets.}

We first define a general typed syntax of interaction nets.

Assume
we are given a set of \emph{symbols} and that an arity (a non-negative integer)
and a typing rule is associated with each symbol. This typing rule is a list
$(A_0,A_1,\dots,A_n)$ of types, where $n$ is the arity associated with the
symbol. Types are formulae of some system of linear logic (in particular if $A$ is a type, $\Orth A$ is also a type and $\BiOrth A=A$). An \emph{interaction net} is made
of \emph{cells}. With each cell $\gamma$ is associated exactly one symbol and
therefore an arity $n$ and a typing rule $(A_0,A_1,\dots,A_n)$. Such a cell
$\gamma$ has one \emph{principal port} $p_0$ and $n$ \emph{auxiliary ports}
$p_1,\dots,p_n$. An interaction net has also a finite set of \emph{free ports}. All these
ports (the free ports and the ports associated with cells) have to be pairwise
distinct and a set of \emph{wires} is given.  This wiring is a set of pairwise
disjoint sets of ports of cardinality $2$ (ordinary wires) or $0$
(loops\footnote{To be more precise, one has to specify the number of loops in
  the net, but this will not play any role in the sequel.}), and the union of
these wires must be equal to the set of all ports of the interaction net. In other words,
each port of the interaction net (free or associated with a cell) is connected to exactly
one other port (free or associated with a cell) through a wire, and each such
wire connects exactly two ports: ports cannot be shared. 
The free ports of the interaction net are those which are not associated with a cell.

An \emph{oriented wire} of the interaction net is an ordered pair $(p_1,p_2)$ where
$\Pair{p_1}{p_2}$ is a wire. In an interaction net, a type is associated with each oriented
wire, in such a way that if $A$ is associated with $(p_1,p_2)$, then $\Orth A$
is associated with $(p_2,p_1)$. Last, the typing rules of the cells must be
respected in the sense that for each cell $\gamma$ of arity $n$, whose ports
are $p_0,p_1,\dots,p_n$ and typing rule is $(A_0,A_1,\dots,A_n)$, denoting by
$p'_0,p'_1,\dots,p'_n$ the ports of the interaction net uniquely defined by the fact that
the sets $\Pair{p_i}{p'_i}$ are wires (for $i=0,1,\dots,n$), then the oriented
wires $(p_0,p'_0)$, $(p'_1,p_1)$,\dots,$(p'_n,p_n)$ have types $A_0$,
$A_1$,\dots,$A_n$ respectively.
The free ports of the interaction net constitute its \emph{interface}. With each
free port $p$ can be associated the type of the unique oriented wire whose
endpoint is $p$: this is the type of $p$ in the interface of the
interaction net. 

Here is a typical example of a typed
interaction net:
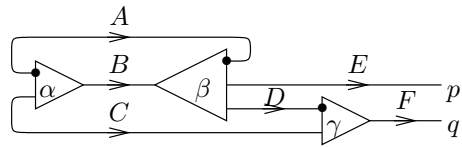
\begin{center}
  \input{int-net-example1.pstex_t}
\end{center}
with cells of symbols $\alpha$, $\beta$ and $\gamma$, of
respective types $(B,\Orth A,\Orth C)$, $(\Orth B,A,\Orth E,\Orth D)$ and
$(F,D,C)$. The interface is $(p:E,q:F)$. Cells are represented as triangles,
with principal port located at one of the angles and other ports on the
opposite edge. We often draw a black dot to locate the auxiliary port
number~$1$.

\paragraph{Dynamics of interaction nets.}

In Lafont's traditional interaction nets, a reduction rule associates, to each pair of cells connected through their principal ports, an interaction net with the same interface as this two-cells net. Moreover at most one rule is given for each pair of cells. The application of a reduction rule substitutes (inside an interaction net) a pair of cells connected through their principal ports by the associated reduct.

These ideas are strongly related with key steps of cut elimination in logics, where two rules interact through their principal formulas.

\paragraph{Extensions of interaction nets.}

Since interaction nets have strong confluence properties, extensions are required to take concurrency and non-determinism into account~\cite{phdalexiev}.

A possible way of extending interaction nets for more concurrent behaviours is to modify statics by introducing cells with many principal ports~\cite{phdalexiev,phdkhalil,Mazza05} (this breaks simple logical interpretations by contradicting the usual property that a logical rule of the sequent calculus introduces \emph{one} principal formula). In differential interaction nets, we stick to the traditional statics syntax (but act on the dynamics side).

A possible extension of the dynamics towards non-deterministic behaviours is to allow a choice between different possible reduction rules for a given pair of cells~\cite{phdalexiev}.

We make a choice in-between this proposal and Lafont's one (\ie{} between many possible rules, and at most one rule). We consider only one reduction rule for each pair of cells but we extend interaction nets to formal sums of interaction nets~\cite{din}. This allows us to represent, as one reduction, a finite set of non-deterministic choices: when the reduct of the interaction of two cells is a (proper) sum of interaction nets.

\subsection{Presentation of the cells}

We define \emph{differential interaction nets} from the interaction nets defined above. We first present our specific cells.

The untyped lambda-calculus can be seen as a typed lambda-calculus equipped with a recursive equation $D=D\Rightarrow D$. Following this idea, V.~Danos and L.~Regnier~\cite{phdregnier} defined a notion of untyped proof-nets 
based on a single type symbol $\Out$
(the type of outputs), subject to the following recursive equation
$\Out=\Par{\Int{\Orth\Out}}\Out$. 
Our nets will be typed using this typing system.
 We set $\In=\Orth\Out$, so that
$\In=\Tens{\Excl\Out}\In$ and $\Out=\Par{\Int\In}\Out$. The tensor connective
is used only with premises $\Excl\Out$ and $\In$ and dually for the par, and
therefore, the only types we actually need are $\Out$, $\In$, $\Excl\Out$ and
$\Int\In$ for typing our nets.

In the present setting, there are ten symbols: par (arity $2$), bottom
(arity $0$), tensor (arity $2$), one (arity $0$), dereliction (arity $1$),
weakening (arity $0$), contraction (arity $2$), codereliction (arity $1$),
coweakening (arity $0$) and cocontraction (arity $2$). We present now the various cell symbols, with their typing rules, in a
pictorial way.

\paragraph{Multiplicative cells.}
The \emph{par} and \emph{tensor} cells, as well as their ``nullary'' versions
\emph{bottom} and \emph{one} are as follows:
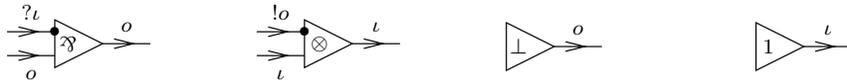
\begin{center}
  \begin{minipage}[c]{0.1\textwidth}\scalebox{1}{\input{par-cell.pstex_t}}\end{minipage}
  \Figspace4
  \begin{minipage}[c]{0.1\textwidth}\scalebox{1}{\input{tensor-cell.pstex_t}}\end{minipage}
  \Figspace4
  \begin{minipage}[c]{0.1\textwidth}\scalebox{1}{\input{bottom-cell.pstex_t}}\end{minipage}
  \Figspace4
  \begin{minipage}[c]{0.1\textwidth}\scalebox{1}{\input{one-cell.pstex_t}}\end{minipage}  
\end{center}

\paragraph{Exponential cells.} They are typed according to a strictly polarized
discipline. Here are first the \emph{why not} cells, which are called
\emph{dereliction}, \emph{weakening} and \emph{contraction}:
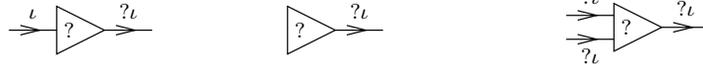
\begin{center}
  \begin{minipage}[c]{0.1\textwidth}\scalebox{1}{\input{dereliction-cell.pstex_t}}\end{minipage}
  \Figspace5
  \begin{minipage}[c]{0.1\textwidth}\scalebox{1}{\input{weakening-cell.pstex_t}}\end{minipage}
  \Figspace5
  \begin{minipage}[c]{0.1\textwidth}\scalebox{1}{\input{contraction-cell.pstex_t}}\end{minipage}
\end{center}
and then the \emph{bang} cells, called \emph{codereliction}, \emph{coweakening}
and \emph{cocontraction}:
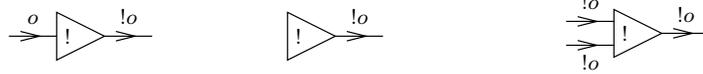
\begin{center}
  \begin{minipage}[c]{0.1\textwidth}\scalebox{1}{\input{codereliction-cell.pstex_t}}\end{minipage}
  \Figspace5
  \begin{minipage}[c]{0.1\textwidth}\scalebox{1}{\input{coweakening-cell.pstex_t}}\end{minipage}
  \Figspace5
  \begin{minipage}[c]{0.1\textwidth}\scalebox{1}{\input{cocontraction-cell.pstex_t}}\end{minipage}
\end{center}

\paragraph{Differential interaction nets.}

We define differential interaction nets from the cells above.

\begin{enumerate}[$\bullet$]
\item
A \emph{simple differential interaction net} is a typed interaction net, which uses the
multiplicative and exponential cells introduced above.
\item A \emph{differential interaction net} is a finite formal sum $S=s_1+\cdots+s_n$ ($n\geq 0$) of
  simple differential interaction nets having all the same interface, and this interface is
  then considered as the interface of $S$. A particular case is the differential interaction net $S=0$
  (the empty sum), and this differential interaction net has to be given together with its interface:
  there is a $0$ net for each interface.
\end{enumerate}

\paragraph{Labeled differential interaction nets.}\label{sec:labnets}
We now introduce labeled differential interaction nets, which are differential interaction nets where particular
cells are equipped with labels. The labeled transition system of differential interaction nets will be defined using these labels in
Section~\ref{sec:transition-system-nets}.

Let $\Labels$ be the countable set of labels obtained by adding a distinguished element
$\NoLab$ (to be understood as the absence of label) to $\LabelsPr$ (introduced in Section~\ref{secsdlts}). A \emph{labeled simple
  differential interaction net} is a simple differential interaction net where all dereliction and codereliction cells
are equipped with labels belonging to $\Labels$. 
Moreover, if $l$ and $l'$ are two labels appearing in a labeled simple differential interaction net, either $l\neq l'$ or $l=l'=\NoLab$.

In our pictures, the labels of dereliction and codereliction cells will
be indicated, when this label is different from $\NoLab$. When its label is
$\NoLab$, a (co)dereliction cell will be drawn without any label.

All the differential interaction nets we consider in this paper are labeled.
In the sequel, since no confusion with other kinds of interaction nets will be
possible, we shall use ``net'' for ``labeled differential interaction net'' and ``simple net'' for ``labeled simple differential interaction net''.

\subsection{Reduction rules}

We denote by $\Nets$ the collection of all simple nets, ranged over by the
letters $s$, $t$, $u$, with or without subscripts or superscripts, and by
$\Fmod\Nat\Nets$ the collection of all nets (finite sums of simple nets with
the same interface, including the empty sum), ranged over by the letters $S$, $T$, $U$, with or without
subscripts or superscripts. We consider $\Nets$ as a subset of
$\Fmod\Nat\Nets$ ($s\in\Nets$ is identified with the sum made of exactly
one copy of $s$).

A \emph{reduction rule} is a subset $\cR$ of
$\Nets\times\Fmod\Nat\Nets$ consisting of pairs $(s,S)$ where $s$ is a
simple net made of two cells connected by their principal ports and
$S$ is a net that has the same interface as $s$. There are actually
reduction rules which transform simple nets into non-simple ones.

This set $\cR$ can be finite or infinite. Such a set is easily extended to
a relation between arbitrary simple nets and nets and then to arbitrary nets and nets:
\begin{enumerate}[$\bullet$]
\item We note $s\Rel\cR T$ if there is $(s_0,u_1+\cdots+u_n)\in\cR$
  where $s_0$ is a subnet of $s$, each $u_i$ is a simple net and
  $T=t_1+\cdots+t_n$ where $t_i$ is the simple net resulting from the
  replacement of $s_0$ by $u_i$ in $s$ ($n\geq 0$).
\item The relation $\Relspan\cR$ is defined by
  $s_1+\cdots+s_n\Rel{\Relspan\cR} T$ (where each $s_i$ is simple) if
  $T=T_1+\cdots+T_n$ where, for each $i$, $s_i\Rel\cR T_i$ or
  $s_i=T_i$ ($n\geq 0$).
\end{enumerate}
Last, the relation $\Reltrans\cR$ between nets and nets is the
transitive closure of $\Relspan\cR$ (which is reflexive).

\subsubsection{Defining the reduction}
We give now the reduction rules of differential interaction nets.

\paragraph{Multiplicative reduction.}
The first two rules concern the interaction of two multiplicative cells of the
same arity.
\begin{center}
  \begin{minipage}[c]{0.3\textwidth}\scalebox{1}{\input{par-tensor-red.pstex_t}}\end{minipage}
  \Figspace5
  \begin{minipage}[c]{0.2\textwidth}\scalebox{1}{\input{bottom-one-red.pstex_t}}\end{minipage}
\end{center}
where $\Emptynet$ stands for the empty simple net (\ie{} the simple net containing no cell, no port and thus no wire, not to be confused with the
net $0\in\Fmod\Nat\Nets$, the empty sum, which is not a simple net). The next
two rules concern the interaction between a binary and a nullary multiplicative
cell.
\begin{center}
  \begin{minipage}[c]{0.35\textwidth}\scalebox{1}{\input{par-one-red.pstex_t}}\end{minipage}
  \Figspace5
  \begin{minipage}[c]{0.35\textwidth}\scalebox{1}{\input{tensor-bottom-red.pstex_t}}\end{minipage}
\end{center}

\paragraph{Communication reduction.} 
Let $R\subseteq\Labels$. We have the following reductions if $l,m\in R$.
\begin{center}
  \begin{minipage}[c]{0.35\textwidth}\scalebox{1}{\input{der-coder-red.pstex_t}}\end{minipage}
\end{center}
The labels $l$ and $m$ disappear in this reduction step.

\paragraph{Non-deterministic reduction.} Let $R\subseteq\Labels$. We have the
following reductions if $l\in R$.
\begin{center}
  \begin{minipage}[c]{0.7\textwidth}\scalebox{1}{\input{der-cocontr-red.pstex_t}}\end{minipage}
  \begin{minipage}[c]{0.7\textwidth}\scalebox{1}{\input{coder-contr-red.pstex_t}}\end{minipage}
\end{center}
\begin{center}
  \begin{minipage}[c]{0.3\textwidth}\scalebox{1}{\input{der-coweak-red.pstex_t}}\end{minipage}
  \quad\quad\quad\quad
  \begin{minipage}[c]{0.3\textwidth}\scalebox{1}{\input{coder-weak-red.pstex_t}}\end{minipage}
\end{center}
According to the definition of reduction for arbitrary simple nets given above, the application of one of the last two steps to a simple net erases the whole simple net and gives the $0$ net (as everywhere else in the paper, this means the $0$ net \emph{with appropriate interface}). These two steps make some labels disappear.

\paragraph{Structural reduction.}
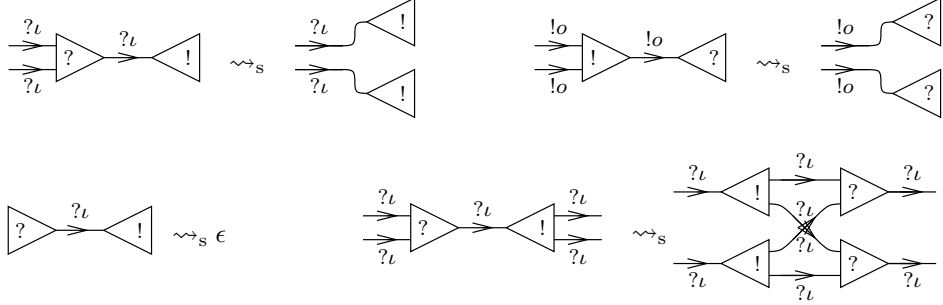
\label{par:struct-red}
\begin{center}
  \begin{minipage}[c]{0.4\textwidth}\scalebox{1}{\input{contr-coweak-red.pstex_t}}\end{minipage}
  \quad\quad
  \begin{minipage}[c]{0.4\textwidth}\scalebox{1}{\input{cocontr-weak-red.pstex_t}}\end{minipage}\\[1em]
  \begin{minipage}[c]{0.2\textwidth}\scalebox{1}{\input{weak-coweak-red.pstex_t}}\end{minipage}
  \quad\quad\quad\quad
  \begin{minipage}[c]{0.55\textwidth}\scalebox{1}{\input{contr-cocontr-red.pstex_t}}\end{minipage}
\end{center}
We use $\NetEqS$ for the symmetric and transitive closure of $\NetRedS$.

\begin{rem}
One can check that we have provided reduction rules for all redexes compatible
with our typing system: for any simple net
$s$ made of two cells connected through their principal ports, there is a
reduction rule whose left member is $s$. This rule is unique, up to the choice
of a set of labels, but this choice has no influence on the right member of the
rule.
\end{rem}

\subsubsection{Confluence}

\begin{thm}\label{thmconfl}
  Let $R,R'\subseteq\Labels$. Let $\cR\subseteq\Nets\times\Fmod\Nat\Nets$ be
  the union of some of the reduction relations $\NetRedCp R$, $\NetRedNDp{R'}$,
  $\NetRedM$ and $\NetRedS$.
  The relation $\Reltrans\cR$ is confluent on $\Fmod\Nat\Nets$.
\qed
\end{thm}

The proof is essentially trivial since the rewriting relation has no critical
pair (see~\cite{din}). Given $R\subseteq\Labels$, we consider in
particular the following \emph{$R$-reduction}: ${\NetRedRel R}=
{\NetRedM}\cup{\NetRedCp{\{\NoLab\}}}\cup{\NetRedS} 
\cup{\NetRedNDp{R}}$.
We set ${\NetRedD}={\NetRedRel\emptyset}$ (``d'' for ``deterministic'') and denote
by $\NetEqD$ the symmetric and transitive closure of this relation.
Observe that, if $s$ and $S$ are nets with $s$ simple and if $s\Rel\NetRedD S$,
then $S$ is also simple.

Some of the reduction rules we have defined depend on a set of labels. This
dependence is clearly monotone in the sense that the relation becomes larger
when the set of labels increases.

The reduction of nets is not normalizing mainly because of the last structural reduction rule. Additional comments can be found in~\cite{din}.

\subsubsection{A transition system of simple nets}
\label{sec:transition-system-nets}

Let $l$ and $m$ be distinct
elements of $\LabelsPr$. We call \emph{$(l,m)$-communication
  redex} a communication redex whose codereliction cell is labeled by $l$ and
whose dereliction cell is labeled by $m$.  

\begin{defi}[The transition system $\NetLTS$]
We define a labeled transition system $\NetLTS$ whose objects are simple nets,
and transitions are labeled by pairs of distinct elements of $\LabelsPr$. Let
$s$ and $t$ be simple nets, we have $s\Rel{\NetTrans lm}t$ if the following
holds: $s\Rel{\Reltrans{\NetRedRel{\{l,m\}}}}s_0+S$ where $s_0$ is
a simple net which contains an $(l,m)$-communication redex 
and becomes $t$ when one reduces this redex.
\end{defi}

\begin{rem}
The non-deterministic steps allowed in the reduction from $s$ to $s_0+S$ can only involve the codereliction and dereliction labeled by $l$ and $m$ respectively. The communication steps only involve $\NoLab$-labeled derelictions and coderelictions. In the solos calculus, solos communicate in one step through a parallel composition. This single step becomes here a sequence of many elementary steps and our restriction allows us to avoid considering the steps which have nothing to do with the communication we are interested in.

  The net $S$ contains other possible communications, so it corresponds to other branches of non-deterministic choices.
\end{rem}

In this paper, to compare transition systems, we only consider \emph{strong} bisimulations~\cite{bisimpark} as given by the following definition.

\begin{defi}[Bisimulation]
  Given two labeled transition systems $\textit{LTS}_1=(S_1,T_1)$ (with $T_1\subseteq S_1\times L\times S_1$) and $\textit{LTS}_2=(S_2,T_2)$ (with $T_2\subseteq S_2\times L\times S_2$) on the same set $L$ of labels, a relation $R$ between $S_1$ and $S_2$ is a \emph{bisimulation} between $\textit{LTS}_1$ and $\textit{LTS}_2$ if:
  \begin{enumerate}[$\bullet$]
  \item for any $(s_1,s_2)\in R$ and $(s_1,l,s'_1)\in T_1$, there exists $s'_2\in S_2$ such that $(s_2,l,s'_2)\in T_2$;
  \item and for any $(s_1,s_2)\in R$ and $(s_2,l,s'_2)\in T_2$, there exists $s'_1\in S_1$ such that $(s_1,l,s'_1)\in T_1$.
  \end{enumerate}
\end{defi}

A useful particular case is $\textit{LTS}_1=\textit{LTS}_2$.

\begin{lem}\label{lemma:neteqd-bisimulation}
  The relation ${\NetEqD}\subseteq\Nets\times\Nets$ is a bisimulation on
  $\NetLTS$.
\qed
\end{lem}

\subsection{A toolbox for process calculi interpretation}\label{sec:toolbox}
We introduce now a few families of simple nets, which are built using the
previously introduced basic cells. They will be used as basic modules for
interpreting processes. All of these nets, but the communication areas, can be
considered as \emph{compound cells}: in reduction, they behave in the same way
as cells of interaction nets.

One could have directly defined these compound cells as primitive constructions for the representation of processes. However we prefer to stress the possibility of implementing these compound objects with more basic ones coming from the logically founded theory of differential linear logic, in order to emphasize the possible application of linear logic tools to the image of our encoding.

Communication areas cannot be considered as cells in our setting since they would require more than one principal port. Interaction nets with many principal ports have been studied (in particular for representing concurrent behaviours~\cite{Mazza05}) but their theory is quite different from the ``one principal port'' case.

\subsubsection{Compound cells}

\paragraph{Generalized contraction and
  cocontraction.}\label{par:generalized-contraction}
A \emph{generalized contraction cell} is a simple
net $\DINstwo$. One of its free ports (which is not connected to an auxiliary port of its cells) is called the principal port of $\DINstwo$. The other free ports (which are finitely many and are not connected to principal ports of cells of the generalized cell) are called the auxiliary ports of $\DINstwo$.
These generalized contraction cells are inductively defined:
\begin{enumerate}[$\bullet$]
\item a wire with type $\Int\In$ is a generalized contraction cell (we select the port with type $\Int\In$ as the principal port of the generalized cell);
\item a weakening cell gives a generalized contraction cell by connecting a wire to its principal port (the only free port of the obtained net is the principal port of the generalized cell);
\item given two generalized contraction cells, by connecting the two auxiliary ports of an additional contraction cell to the principal ports of the generalized cells and by connecting the principal port of this new contraction cell to a free port, one obtains a generalized contraction cell (the free port connected to the principal port of the added contraction cell is the principal port of the generalized cell).
\end{enumerate}
Generalized cocontraction cells are defined dually.

We use the same graphical notations for generalized (co)contraction cells as
for ordinary (co)contraction cells, with a ``$*$'' in superscript to the
``$!$'' or ``$?$'' symbols to avoid confusions. Observe that there are
infinitely many generalized (co)contraction cells of any given arity.
Even if one could define a choice of one \emph{canonical} (co)contraction cell for each arity, pluging two canonical (co)contraction cells together would not always give us a canonical one. This is why we do not define such a choice here.

\begin{figure}
  \centering
  \begin{minipage}[c]{0.6\textwidth}\scalebox{1}{\input{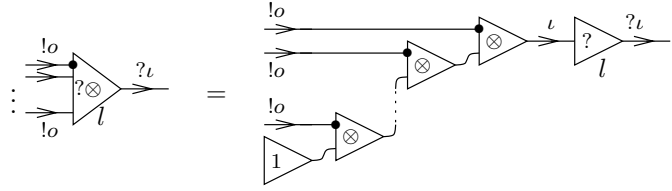}}\end{minipage}
  \caption{Dereliction-tensor compound cell.}
  \label{fig:dereliction-tensor}
\end{figure}

\paragraph{The dereliction-tensor and the codereliction-par cells.} Let $n$ be
a non-negative integer. We define an $n$-ary $\ITensexp$ compound cell as in
Figure~\ref{fig:dereliction-tensor}. It will be decorated by the label of its
dereliction cell (as mentioned in Section~\ref{sec:labnets}, we put no decoration if the label is $\NoLab$).
The number of tensor cells in this compound cell is equal to $n$ (in particular the $0$-ary dereliction-tensor cell contains exactly two cells: a one cell and a dereliction cell). We define dually the $\IParexp$ compound cell.

\paragraph{The prefix cells.} Now we can define the compound cells which will
play the main role in the interpretation of solos. Thanks to the above defined cells, all the oriented wires of
the nets we shall define will have type $\Int\In$ or $\Excl\Out$. Therefore, we
adopt the following graphical convention: the wires will have an orientation
corresponding to the $\Int\In$ type.

\begin{figure}
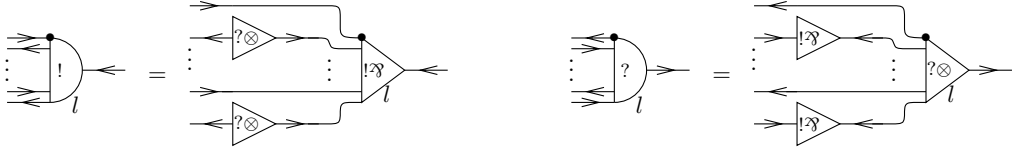

  \centering
  \begin{minipage}[c]{0.45\textwidth}\scalebox{0.9}{\input{input-cell-def.pstex_t}}\end{minipage}
  \Figspace1
  \begin{minipage}[c]{0.45\textwidth}\scalebox{0.9}{\input{output-cell-def.pstex_t}}\end{minipage}
  \caption{Input and output compound cells.}
  \label{fig:input-output-cell}
\end{figure}

Let $n$ be a non-negative integer.
The \emph{$n$-ary input cell} and the \emph{$n$-ary output cell} are defined in
Figure~\ref{fig:input-output-cell}, they have $n$ pairs of auxiliary ports $(\delta^+_1,\delta^-_1,\dots,\delta^+_n,\delta^-_n)$.

The label of a prefix cells is the one carried by its outermost $\ITensexp$ or
$\IParexp$ compound cell, the other $\ITensexp$ or
$\IParexp$ compound cells are required to be unlabeled (that is, labeled by $\NoLab$).

\subsubsection{Communication areas}
\label{seccommarea}

\begin{figure}
\centering
\scalebox{1}{\input{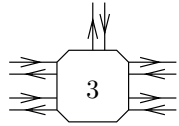}}
\caption{Area of order 3.}\label{fig:comm3}
\end{figure}
Let $n\geq -2$. We define a family of nets with $2(n+2)$ free ports, called
\emph{communication areas of order $n$}, that we shall draw using rectangles
with beveled angles. Figure~\ref{fig:comm3} shows how we picture a
communication area of order $3$.

A communication area of order $n$ is made of $n+2$ pairs of $(n+1)$-ary
generalized cocontraction and contraction cells
$(\gamma^+_1,\gamma^-_1),\dots,(\gamma^+_{n+2},\gamma^-_{n+2})$, with, for each
$i$ and $j$ such that $1\leq i<j\leq n+2$, a wire from an auxiliary port of
$\gamma^+_i$ to an auxiliary port of $\gamma^-_j$ and a wire from an auxiliary
port of $\gamma^-_i$ to an auxiliary port of $\gamma^+_j$.
We note $p^+_i$ and $p^-_i$ the principal ports of $\gamma^+_i$ and $\gamma^-_i$, and we call ($p^+_i,p^-_i$) a pair of \emph{associated ports} of the communication area.

\begin{figure}
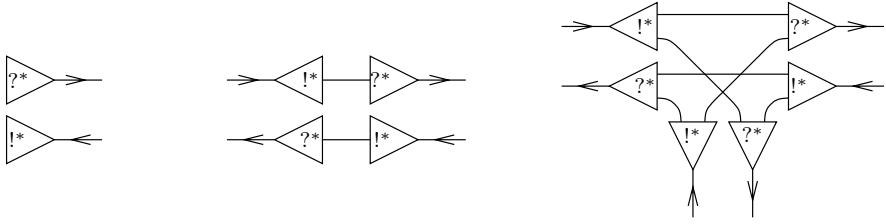

  \centering
  \begin{minipage}[c]{0.15\textwidth}\scalebox{1}{\input{comm-area--1.pstex_t}}\end{minipage}
  \Figspace1
  \begin{minipage}[c]{0.25\textwidth}\scalebox{1}{\input{comm-area-0.pstex_t}}\end{minipage}
  \Figspace1
  \begin{minipage}[c]{0.35\textwidth}\scalebox{1}{\input{comm-area-1.pstex_t}}\end{minipage}
  \caption{Communication areas of order $-1$, $0$ and $1$.}
  \label{fig:communication-examples}
\end{figure}

So the communication area of order $-2$ is the empty net $\Emptynet$, and
communication areas of order $-1$, $0$ and $1$ are the structures shown in
Figure~\ref{fig:communication-examples}.

\subsubsection{Useful reductions.}

One of the nice properties of communication areas is that, when one connects
two such areas through a pair of wires, one gets another communication area.

\begin{figure}
  \centering
  \scalebox{1}{\input{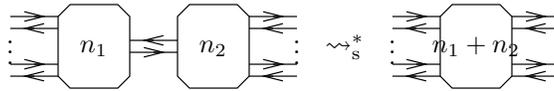}}
  \caption{Aggregation, with $n_1,n_2\geq -1$.}\label{fig:aggregation}
\end{figure}

\begin{lem}[Aggregation of communication areas]\label{lemaggreg}
  Let $C_1$ be a communication area of order $n_1\geq -1$ and $C_2$ be a communication area of order $n_2\geq -1$, let $(p^+_1,p^-_1)$ be a pair of associated ports of $C_1$ and $(p^+_2,p^-_2)$ be a pair of associated ports of $C_2$, let $\DINsone$ be the simple net obtained by connecting $p^+_1$ to $p^-_2$ and $p^-_1$ to $p^+_2$, we have $\DINsone\Reltrans\NetRedS C$ where $C$ is a communication area of order $n_1+n_2$ (see Figure~\ref{fig:aggregation}).
\end{lem}

\proof
  We only prove the particular case where $n_1=n_2=1$ and the communication areas $C_1$ and $C_2$ are built with contraction cells (not generalized ones).

\begin{figure}
  \centering
    \scalebox{0.8}{\input{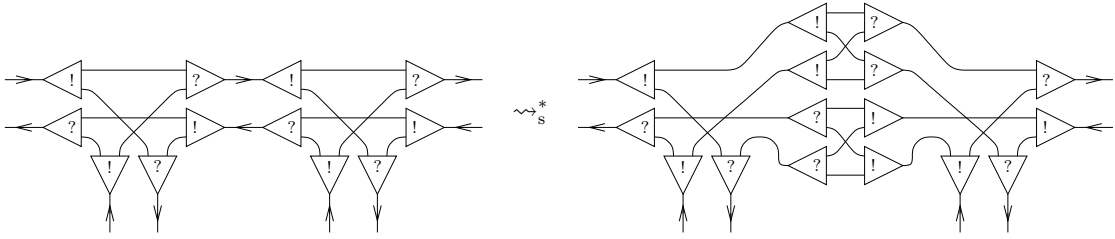}}
  \caption{Aggregation of communication areas of order $1$.}
  \label{fig:proofaggreg}
\end{figure}

  By connecting $C_1$ and $C_2$, and by applying two structural reduction steps, we obtain a communication area of order $2$ (see Figure~\ref{fig:proofaggreg}).
\qed

\begin{figure}
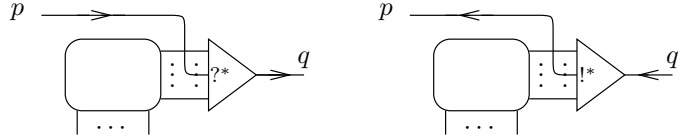

  \centering
  \scalebox{1}{\input{der-forwarding.pstex_t}}
  \Figspace2
  \scalebox{1}{\input{coder-forwarding.pstex_t}}
  \caption{Port forwarding.}
  \label{fig:port-forwarding}
\end{figure}

Let $\DINstwo$ be a simple net and $p$ be a 
port of $\DINstwo$. We say that \emph{$p$ is
  forwarded in $\DINstwo$} if there is a free port $q$ of $\DINstwo$ such that $\DINstwo$ is of one
of the two shapes given in Figure~\ref{fig:port-forwarding}.

This allows us to describe the interaction of derelictions and coderelictions with communication areas: derelictions and coderelictions can meet each other, when connected to a common communication area.

\begin{lem}[Communication and forwarding of derelictions and coderelictions in communication areas]\label{lemdercoder}
Let $p\geq -2$ be an integer, let $l,m\in\Labels$, and let us consider the simple net $\DINsone$ obtained by connecting (the principal ports of) a codereliction labeled $l$ and a weakening to a pair of associated ports of a communication area of order $p+2$, and by connecting (the principal ports of) a dereliction labeled $m$ and a coweakening to another pair of associated ports of the same communication area. The $\{l,m\}$-reduction applied to $\DINsone$ leads to a sum of simple nets: one contains the codereliction $l$ and the dereliction $m$ connected through their principal ports, together with a communication area of order $p$; in all the other summands, the principal ports $r$ and $r'$ of the codereliction $l$ and of the dereliction $m$ are forwarded (see Figure~\ref{fig:der-coder-communication}).

\begin{figure}
  \centering
  \scalebox{1}{\input{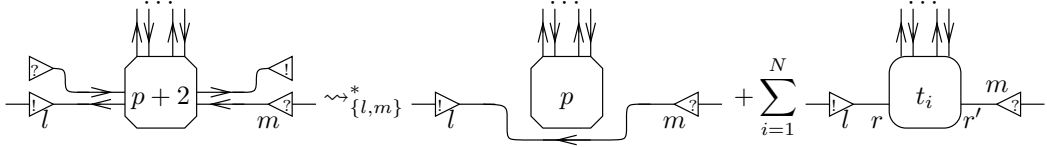}}
  \caption{Dereliction and codereliction communicating through a communication
    area.}\label{fig:der-coder-communication}
\end{figure}
\end{lem}

\begin{figure}
  \centering
  \scalebox{0.6}{\input{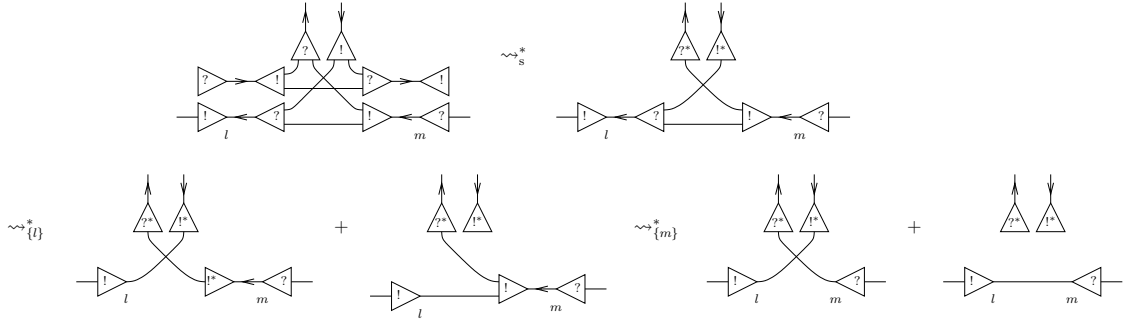}}
  \caption{Forwarding of dereliction and codereliction.}\label{fig:proofcommforward}
\end{figure}

\proof
  We consider the particular case where $p=-1$ and the communication area is built with contraction cells (not generalized ones).
  The reduction is pictured in Figure~\ref{fig:proofcommforward}.

  We first apply all the possible structural reduction steps, then we focus on the codereliction labeled $l$: the reduction of the associated redex gives a sum of two simple nets (in one of them the principal port of the codereliction $l$ is forwarded). We now apply in each simple net the reduction step involving the dereliction labeled $m$ and we finally obtain a sum of two simple nets. In the first one the principal ports of the dereliction and of the codereliction are forwarded, in the second one they are connected to each other and the remaining part of the net is a communication area of order $-1$.

 The generalization to other communication areas of order $1$ is easy. We can obtain the case $p\neq -1$ by decomposing the communication area into two communication areas (one of order $1$ and one of order $p+1$). Lemma~\ref{lemaggreg} and Theorem~\ref{thmconfl} help to conclude.
\qed

We now consider the interaction of two prefix cells.

\begin{lem}[Reduction of prefixes]\label{lemprefix}
Let $l,m\in\Labels$.  If we connect an $n$-ary output prefix labeled by $m$ to
a $p$-ary input prefix labeled by $l$ through their principal ports, we obtain a simple net which reduces by
$\NetRedCp{\{l,m\}}$ to a net $u$ which reduces by
$\Reltrans{\NetRedRel{\{\NoLab\}}}$ to $0$ if $n\not=p$ and to simple wires by
$\Reltrans{\NetRedRel\emptyset}$, as in
Figure~\ref{fig:in-out-prefix-reduction}, if $n=p$.

\begin{figure}
  \centering
    \scalebox{1}{\input{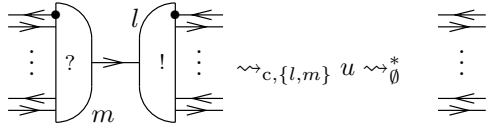}}
  \caption{Prefix reduction.}
  \label{fig:in-out-prefix-reduction}
\end{figure}
\end{lem}

\proof
  The key point is to check that the simple net, obtained by connecting an $n$-ary dereliction-tensor compound cell to a $p$-ary codereliction-par cell through their principal ports, reduces to $\min(n,p)$ wires with $n-p$ weakening cells if $n>p$ or $p-n$ coweakening cells if $p>n$.
\qed

\section{From solo diagrams to differential interaction nets}\label{sectrans}

\subsection{The translation}

Relying on the toolbox described above, we define a translation of labeled solo diagrams with identifications into labeled simple differential interaction nets:
\begin{enumerate}[$\bullet$]
\item A node which appears $n$ times as a source of a multiedge, which is $k$ times target of a multiedge and which is $p$ times member of an identification edge is translated as a \emph{communication area} of order $n+k+p-2$.

For example, for a node which is target of one input multiedge and of one output multiedge (thus $k=2$), which is source of one multiedge (thus $n=1$), and which is member of one identification edge (thus $p=1$), we have $n+k+p-2=2$:
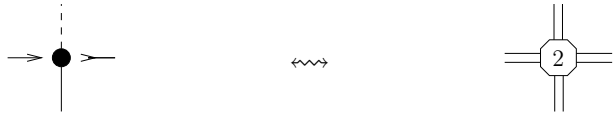
\begin{center}
\begin{tabular}[c]{c}
\scalebox{0.75}{\begin{picture}(0,0)%
\includegraphics{trad-solo-node.pstex}%
\end{picture}%
\setlength{\unitlength}{3947sp}%
\begingroup\makeatletter\ifx\SetFigFontNFSS\undefined%
\gdef\SetFigFontNFSS#1#2#3#4#5{%
  \reset@font\fontsize{#1}{#2pt}%
  \fontfamily{#3}\fontseries{#4}\fontshape{#5}%
  \selectfont}%
\fi\endgroup%
\begin{picture}(924,924)(1039,-523)
\end{picture}%
}
\end{tabular}
\hspace{2cm}$\SDDINRel$\hspace{2cm}
\begin{tabular}[c]{c}
\scalebox{0.75}{\begin{picture}(0,0)%
\includegraphics{trad-din-node.pstex}%
\end{picture}%
\setlength{\unitlength}{3947sp}%
\begingroup\makeatletter\ifx\SetFigFontNFSS\undefined%
\gdef\SetFigFontNFSS#1#2#3#4#5{%
  \reset@font\fontsize{#1}{#2pt}%
  \fontfamily{#3}\fontseries{#4}\fontshape{#5}%
  \selectfont}%
\fi\endgroup%
\begin{picture}(924,925)(1039,-935)
\put(1501,-543){\makebox(0,0)[b]{\smash{{\SetFigFontNFSS{12}{14.4}{\rmdefault}{\mddefault}{\updefault}$2$}}}}
\end{picture}%
}
\end{tabular}
\end{center}
\vspace{1ex}
\item An input multiedge with sources $[\node{x}_1,\node{x}_2,\node{x}_3]$ and target $\node{u}$ is translated as a $3$-ary \emph{input cell} with principal port $u^0$ and auxiliary ports $[x_1^0,x_1^1,x_2^0,x_2^1,x_3^0,x_3^1]$. Each pair $x_i^0$, $x_i^1$ is connected to a pair of associated ports of the communication area corresponding to $\node{x}_i$. $u^0$ is connected to the port $p^-$ and a weakening cell is connected to $p^+$, where $(p^+,p^-)$ is a pair of associated ports of the communication area corresponding to $\node{u}$. The label of the prefix cell is the label of the multiedge.
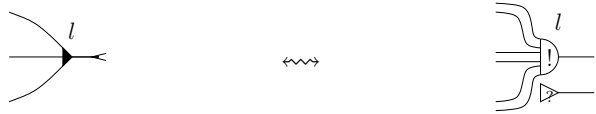
\begin{center}
\begin{tabular}[c]{c}
\scalebox{0.75}{\begin{picture}(0,0)%
\includegraphics{trad-solo-input.pstex}%
\end{picture}%
\setlength{\unitlength}{3947sp}%
\begingroup\makeatletter\ifx\SetFigFontNFSS\undefined%
\gdef\SetFigFontNFSS#1#2#3#4#5{%
  \reset@font\fontsize{#1}{#2pt}%
  \fontfamily{#3}\fontseries{#4}\fontshape{#5}%
  \selectfont}%
\fi\endgroup%
\begin{picture}(836,774)(664,-448)
\put(1201, 89){\makebox(0,0)[b]{\smash{{\SetFigFontNFSS{12}{14.4}{\rmdefault}{\mddefault}{\updefault}$l$}}}}
\end{picture}%
}
\end{tabular}
\hspace{2cm}$\SDDINRel$\hspace{2cm}
\begin{tabular}[c]{c}
\scalebox{0.75}{\input{trad-din-input.pstex_t}}
\end{tabular}
\end{center}
\vspace{1ex}
\item An output multiedge with sources $[\node{x}_1,\node{x}_2,\node{x}_3]$ and target $\node{u}$ is translated as a $3$-ary \emph{output cell} with principal port $u^0$ and auxiliary ports $[x_1^0,x_1^1,x_2^0,x_2^1,x_3^0,x_3^1]$. Each pair $x_i^0$, $x_i^1$ is connected to a pair of associated ports of the communication area corresponding to $\node{x}_i$. $u^0$ is connected to the port $p^+$ and a coweakening cell is connected to $p^-$, where $(p^+,p^-)$ is a pair of associated ports of the communication area corresponding to $\node{u}$. The label of the prefix cell is the label of the multiedge.
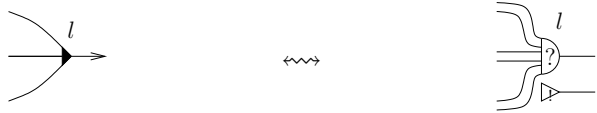
\begin{center}
\begin{tabular}[c]{c}
\scalebox{0.75}{\begin{picture}(0,0)%
\includegraphics{trad-solo-output.pstex}%
\end{picture}%
\setlength{\unitlength}{3947sp}%
\begingroup\makeatletter\ifx\SetFigFontNFSS\undefined%
\gdef\SetFigFontNFSS#1#2#3#4#5{%
  \reset@font\fontsize{#1}{#2pt}%
  \fontfamily{#3}\fontseries{#4}\fontshape{#5}%
  \selectfont}%
\fi\endgroup%
\begin{picture}(849,774)(514,-448)
\put(1051, 89){\makebox(0,0)[b]{\smash{{\SetFigFontNFSS{12}{14.4}{\rmdefault}{\mddefault}{\updefault}$l$}}}}
\end{picture}%
}
\end{tabular}
\hspace{2cm}$\SDDINRel$\hspace{2cm}
\begin{tabular}[c]{c}
\scalebox{0.75}{\input{trad-din-output.pstex_t}}
\end{tabular}
\end{center}
\vspace{1ex}
\item An identification edge connecting $\node{x}_1$ and $\node{x}_2$ is translated as a pair of wires connecting a pair $(p_1^+,p_1^-)$ of associated ports of the communication area corresponding to $\node{x}_1$ with a pair $(p_2^-,p_2^+)$ of associated ports of the communication area corresponding to $\node{x}_2$.
\begin{center}
\begin{tabular}[c]{c}
\scalebox{0.75}{\begin{picture}(0,0)%
\includegraphics{trad-solo-ident.pstex}%
\end{picture}%
\setlength{\unitlength}{3947sp}%
\begingroup\makeatletter\ifx\SetFigFontNFSS\undefined%
\gdef\SetFigFontNFSS#1#2#3#4#5{%
  \reset@font\fontsize{#1}{#2pt}%
  \fontfamily{#3}\fontseries{#4}\fontshape{#5}%
  \selectfont}%
\fi\endgroup%
\begin{picture}(1066,166)(368,-144)
\end{picture}%
}
\end{tabular}
\hspace{2cm}$\SDDINRel$\hspace{2cm}
\begin{tabular}[c]{c}
\scalebox{0.75}{\begin{picture}(0,0)%
\includegraphics{trad-din-ident.pstex}%
\end{picture}%
\setlength{\unitlength}{3947sp}%
\begingroup\makeatletter\ifx\SetFigFontNFSS\undefined%
\gdef\SetFigFontNFSS#1#2#3#4#5{%
  \reset@font\fontsize{#1}{#2pt}%
  \fontfamily{#3}\fontseries{#4}\fontshape{#5}%
  \selectfont}%
\fi\endgroup%
\begin{picture}(1224,324)(289,-223)
\end{picture}%
}
\end{tabular}
\end{center}
\end{enumerate}

\noindent Since communication areas of order $n$ are not uniquely defined (because generalized contraction and cocontraction cells of a given arity are not unique either), this translation from solo diagrams to differential interaction nets is in fact a relation, which we denote $\SDone\SDDINRel\DINsone$ (with $\SDone$ a solo diagram and $\DINsone$ a simple differential interaction net).

\begin{rem}\label{remnormd}
Let $\SDone$ be a solo diagram (without identification edges) and $\DINsone$ be a differential interaction net, if $\SDone\SDDINRel\DINsone$ then $\DINsone$ has no $\NetRedD$ redex, except some $\NetRedS$ redexes involving weakening or coweakening cells. Indeed, all the redexes in $\DINsone$ are given by a labeled codereliction cell (\ie{} whose label is not $\NoLab$) or a coweakening cell facing a weakening cell, a contraction cell or a labeled dereliction cell (or dually by a dereliction cell or a weakening cell facing a coweakening or ...).
\end{rem}

\begin{figure}
    \centering
\scalebox{0.75}{\input{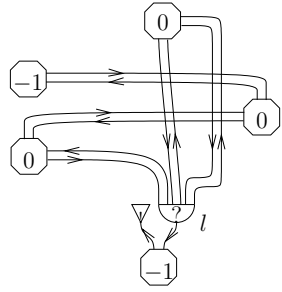}}
    \caption{The translation of the second solo diagram of Figure~\ref{figexsdid}.}\label{figdinex1}
  \end{figure}

A differential interaction net associated with the second solo diagram with identifications of Figure~\ref{figexsdid} is given in Figure~\ref{figdinex1}.

\subsection{A bisimulation}

Our goal is to establish a bisimulation between the labeled transition systems $\SDLTS$ (Section~\ref{secsdlts}) and $\NetLTS$ (Section~\ref{sec:transition-system-nets}). Unfortunately the translation $\SDDINRel$ does not provide such a bisimulation. We are going to show what the problems are and how to restrict $\SDLTS$ to get a bisimulation.

\subsubsection{Mismatch}

The crucial point comes from step~(\ref{stepcontract}) of the reduction of solo diagrams: the contraction of identification edges in solo diagrams corresponds through $\SDDINRel$ to the aggregation of communication areas as given by the following lemma.

\begin{lem}\label{lemredident}
If $\SDone_1$ and $\SDone_2$ are solo diagrams with identifications, if $\SDone_2$ is obtained from $\SDone_1$ by contracting an identification edge (step~(\ref{stepcontract})) connecting the nodes $\node{n}_1$ and $\node{n}_2$ with $\node{n}_1\neq \node{n}_2$ and if $\SDone_1\SDDINRel\DINsone_1$ then $\SDone_2\SDDINRel\DINsone_2$ where $\DINsone_2$ is obtained from $\DINsone_1$ by aggregating the communication areas corresponding to $\node{n}_1$ and $\node{n}_2$.
\end{lem}

\proof
  \begin{figure}
    \centering
    \scalebox{0.75}{\input{proofredident.pstex_t}}
    \caption{Proof of Lemma~\ref{lemredident}.}\label{figlemredident}
  \end{figure}
See Figure~\ref{figlemredident}:
$\SDone_2$ is obtained from $\SDone_1$ by replacing the two distinct nodes $\node{n}_1$ and $\node{n}_2$ by a unique node $\node{n}$. In $\DINsone_1$, we have a communication area $C_1$ corresponding to $\node{n}_1$ and a communication area $C_2$ corresponding to $\node{n}_2$. Since $\node{n}_1$ and $\node{n}_2$ are connected with an identification edge, $C_1$ and $C_2$ are connected with a pair of wires. By aggregating $C_1$ and $C_2$ (see Lemma~\ref{lemaggreg}), we obtain a communication area $C$ in $\DINsone_2$ which exactly corresponds to $\node{n}$ through $\SDDINRel$.
\qed

The hypothesis $\node{n}_1\neq \node{n}_2$ is crucial since two communication areas $C_1$ and $C_2$ connected together with a pair of wires reduce to one communication area (Lemma~\ref{lemaggreg}), \emph{except} if $C_1=C_2$! We are thus able to encode the reduction of the solos calculus only if we can ensure that we never have to contract an identification edge connecting a node with itself.

A typical example of this problem is given by:
\begin{equation*}
\SoloNu u{\SoloNu x{\SoloNu y{\SoloNu z{\SoloNu{y'}{\SoloNu{z'}{\SoloNu{a}{\SoloNu{b}{\SoloNu{c}{\SoloNu{a'}{\SoloNu{b'}{\SoloNu{c'}{\;(\SoloPar{\SoloPar{\SoloOut u{xyz}}{\SoloIn u{xy'z'}}}{\SoloPar{\SoloOut x{abc}}{\SoloIn x{a'b'c'}}})}}}}}}}}}}}}
\end{equation*}
As shown in Figure~\ref{fig:redcomp}, while in solo diagrams the loop created after the first step eventually disappears, it remains in differential interaction nets. The consequence being that we have transitions in $\SDLTS$ while the last differential interaction net has no transition
to a differential interaction net which is the translation of a solo diagram.

\begin{figure}
\centering
\begin{tabular}{ccc}
\begin{tabular}[c]{c}
\scalebox{0.75}{\input{soloac1bis.pstex_t}}
\end{tabular}
& \hspace{4cm} & 
\begin{tabular}[c]{c}
\scalebox{0.75}{\input{dinac1bis.pstex_t}}
\end{tabular}
\\
\\
$\xrightarrow{\text{(\ref{stepident})}}$ & & $\NetRedRel{\{l,m\}}^+$ \\
\\
\begin{tabular}[c]{c}
\scalebox{0.75}{\input{soloac2bis.pstex_t}}
\end{tabular}
& & 
\begin{tabular}[c]{c}
\scalebox{0.75}{\input{dinac2bis.pstex_t}}
\end{tabular}
\\
\\
$\xrightarrow{\text{(\ref{stepcontract})}^2}$ & & $\NetRedD^+$ \\
\\
\begin{tabular}[c]{c}
\scalebox{0.75}{\input{soloac3bis.pstex_t}}
\end{tabular}
& & 
\begin{tabular}[c]{c}
\scalebox{0.75}{\input{dinac3bis.pstex_t}}
\end{tabular}
\\
$\xrightarrow{\text{(\ref{stepcontract})}}$ \\
\\
\begin{tabular}[c]{c}
\scalebox{0.75}{\input{soloac4bis.pstex_t}}
\end{tabular}
\\
\\
$\xrightarrow{\text{(\ref{stepident})}}$ \\
\\
\begin{tabular}[c]{c}
\scalebox{0.75}{\input{soloac5bis.pstex_t}}
\end{tabular}
\\
\\
$\xrightarrow{\text{(\ref{stepcontract})}^3}$ \\
\\
\begin{tabular}[c]{c}
$\emptyset$
\end{tabular}
\end{tabular}
\caption{Reductions in solo diagrams and in differential interaction nets, corresponding to the term:\hfill\break $\SoloNu u{\SoloNu x{\SoloNu y{\SoloNu z{\SoloNu{y'}{\SoloNu{z'}{\SoloNu{a}{\SoloNu{b}{\SoloNu{c}{\SoloNu{a'}{\SoloNu{b'}{\SoloNu{c'}{\;(\SoloPar{\SoloPar{\SoloOut u{xyz}}{\SoloIn u{xy'z'}}}{\SoloPar{\SoloOut x{abc}}{\SoloIn x{a'b'c'}}})}}}}}}}}}}}}$.}\label{fig:redcomp}
\end{figure}

\subsubsection{Restriction}
\label{secrestrlts}

Since our goal is to make the bisimulation result true, we are going to constrain the syntax of solo diagrams in such a way that the hypotheses of Lemma~\ref{lemredident} are always valid. 
We want to restrict the reduction step~(\ref{stepcontract}) to the case where the edge is not connecting a node with itself. This is equivalent to asking the sub-graph $\NdSDred{\SDone}{e_1,e_2}$ of $\SDred{\SDone}{e_1,e_2}$, containing the identification edges only, to be acyclic.
By definition, we call \emph{acyclic reduction step} a reduction step associated with two dual multiedges $e_1$ and $e_2$ such that $\NdSDred{\SDone}{e_1,e_2}$ is acyclic.

Another problem comes from the constraint on the freeness of names in the contraction of identification edges in solo diagrams. Our translation into differential interaction nets is forgetting the fact that some nodes are free and others are bound. As a consequence a reduction might happen in the differential interaction net's side without being possible in the solo diagram's side. An example is given by the term $\SoloPar{\SoloOut uxyz}{\SoloIn ux'y'z'}$. To avoid this situation we introduce the notion of \emph{acyclic redex}.

\begin{defi}[Acyclic redex]\label{defacredex}
In a solo diagram, a pair of dual multiedges is an \emph{acyclic redex} if it is a redex (meaning that the induced freeness conditions are satisfied) and the induced reduction step is an acyclic reduction step.
\end{defi}
 
\begin{defi}[$\SDLTSac$ labeled transition system]\label{defltssdac}
$\SDLTSac$ is the biggest labeled transition system such that:
\begin{enumerate}[$\bullet$]
\item each object of $\SDLTSac$ is an object of $\SDLTS$ and each transition of $\SDLTSac$ is a transition of $\SDLTS$ (but some objects and transitions are lost),
\item for any object in $\SDLTSac$, all the transitions starting from it in $\SDLTS$ belong to $\SDLTSac$,
\item all pairs of dual multiedges in objects of $\SDLTSac$ are acyclic redexes.
\end{enumerate}
\end{defi}

\noindent This means that an object $G$ in $\SDLTS$ belongs to $\SDLTSac$ as soon as none of the paths starting from $G$ in $\SDLTS$ allows us to reach an object with a non-acyclic redex. This is a decidable property since there are finitely many such paths.

\begin{defi}[Acyclic solo diagram]\label{defacdiag}
A solo diagram which is an object of $\SDLTSac$ is called an \emph{acyclic solo diagram}.
\end{defi}

\begin{prop}\label{propbisimone}
If $\SDone\SDTrans lm\SDtwo$ in $\SDLTSac$ and $\SDone\SDDINRel\DINsone$, then there are simple nets $\DINstwo$ and $\DINstwo_0$ such that $\DINsone\NetTrans lm\DINstwo_0$ in $\NetLTS$, $\SDtwo\SDDINRel\DINstwo$ and $\DINstwo_0\NetRedD\DINstwo$.
\end{prop}

\proof
By hypothesis, there is a communication area $C$ in $\DINsone$ with two dual prefixes labeled with $l$ and $m$ connected to it. We apply the forwarding of derelictions and coderelictions in the communication area $C$ to them (Lemma~\ref{lemdercoder}). This gives us a sum of simple nets containing a simple net $\DINsone_1$ with an $(l,m)$-communication redex and other summands where $l$ and $m$ have been forwarded.

\begin{figure}
  \centering
  \scalebox{0.75}{\input{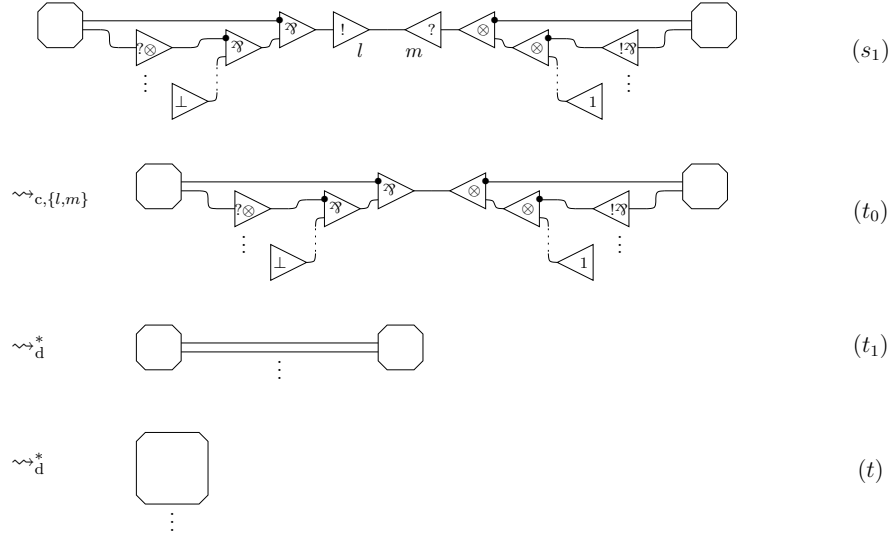}}
  \caption{Simulation of a solos reduction step.}\label{figproofpropbisimone}
\end{figure}
Following Figure~\ref{figproofpropbisimone},
let $\DINstwo_0$ be the simple net obtained by reducing the $(l,m)$-communication redex of $\DINsone_1$, and $\DINstwo_1$ be the simple net obtained by applying to $\DINsone_1$ the prefix reduction starting with the reduction from $\DINsone_1$ to $\DINstwo_0$ (see Lemma~\ref{lemprefix}). It is easy to see that $\SDred{\SDone}{e_1,e_2}\SDDINRel\DINstwo_1$ where $e_1$ and $e_2$ are the multiedges of $\SDone$ with labels $l$ and $m$.
By Lemma~\ref{lemredident}, each step of acyclic contraction of an identification edge corresponds to the aggregation of the two associated communication areas, thus $\SDtwo\SDDINRel\DINstwo$ with $\DINstwo_1\Reltrans\NetRedD\DINstwo$ and we have $\DINstwo_0\Reltrans\NetRedD\DINstwo$.
\qed

\begin{lem}[Diving]\label{lemdiving}
Let $\SDone$ be a solo diagram (without identification), if $\SDone\SDDINRel\DINsone$ and $\DINsone\Reltrans{\NetRedRel{\{l,m\}}}\DINstwo+\DINtwo$ with $\DINstwo$ containing an $(l,m)$-communication redex, then the codereliction labeled $l$ and the dereliction labeled $m$ are connected to the same communication area $C$ in $\DINsone$ and the $(l,m)$-communication redex is generated by the dereliction/codereliction forwarding through $C$ (Lemma~\ref{lemdercoder}).
\end{lem}

\proof
According to Remark~\ref{remnormd}, the only $\NetRedRel{\{l,m\}}$ redexes of $\DINsone$ are involving the codereliction $l$, the dereliction $m$, a weakening or a coweakening.

\begin{figure}
  \centering
  \scalebox{0.75}{\input{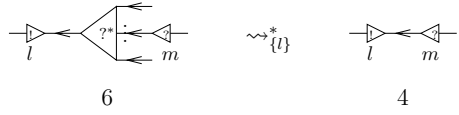}}
  \caption{Proof of Lemma~\ref{lemdiving}: sequence~\ref{caselpcontr}$^*$-\ref{caselm}.}
  \label{figproofdiving1}
\end{figure}

\begin{figure}
  \centering
  \scalebox{0.75}{\input{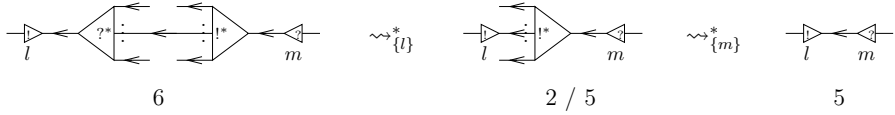}}
  \caption{Proof of Lemma~\ref{lemdiving}: sequence~\ref{caselpcontr}$^*$-\ref{caselauxcontr}-\ref{casempcontr}$^+$.}
  \label{figproofdiving2}
\end{figure}

By confluence (Theorem~\ref{thmconfl}), we can assume the reductions involving the codereliction $l$ are coming before those involving $m$ (and those involving the weakeaning and coweakening cells arrive after them). We look at the port connected to the principal port of the codereliction $l$:
\begin{enumerate}[($l1$)]
\item If it is the auxiliary port of a prefix cell, there is no reduction for $l$, and no $\{l,m\}$-reduction will change the situation (or this prefix cell is labeled $m$ and changing the situation means erasing $m$).
\item\label{caselauxcontr} If it is the auxiliary port of a generalized cocontraction cell (which is not a wire), there is no reduction for $l$.
\item If it is the principal port of a prefix cell labeled $k$ ($k\neq m$), there is no $\{l,m\}$-reduction for $l$, and no $\{l,m\}$-reduction will change the situation.
\item\label{caselm} If it is the principal port of the prefix cell labeled $m$, we immediately have the result, and $C$ is a communication area of order $0$ (with the part connecting $l$ and $m$ which is just a simple wire).
\item If it is the principal port of a generalized contraction cell of arity $0$, reductions involving $l$ will eventually erase it and we will never reach an $(l,m)$-communication redex, this contradicts the hypotheses.
\item\label{caselpcontr} If it is the principal port of a generalized contraction cell of arity $n>0$, in order to reach an $(l,m)$-communication redex, the reductions involving $l$ must start destroying the generalized contraction.
In order to eventually reach an $(l,m)$-communication redex, we must have a simple net corresponding to case~\ref{caselpcontr} again, or to case~\ref{caselauxcontr} or~\ref{caselm}.
\end{enumerate}
We now consider these last two configurations: \ref{caselpcontr} cases followed by~\ref{caselauxcontr} (see Figure~\ref{figproofdiving2}) or~\ref{caselm} (see Figure~\ref{figproofdiving1}).
Note that we cannot have infinitely many consecutive \ref{caselpcontr} cases since the induced reduction makes the size of the net decrease. If after \ref{caselpcontr} cases we reach case~\ref{caselm}, $l$ and $m$ where connected to the same communication area $C$ and we have the result (\ref{caselpcontr} cases give a forwarding of $l$ through $C$).
If after \ref{caselpcontr} cases we reach case~\ref{caselauxcontr}, we turn our attention to the reduction steps involving $m$. We look at the port connected to the principal port of the dereliction $m$:
\begin{enumerate}[($m1$)]
\item If it is the auxiliary port of a prefix cell, there is no reduction for $m$, and no $\{l,m\}$-reduction will change the situation (or this prefix cell is labeled $l$ and changing the situation means erasing $l$).
\item If it is the auxiliary port of a generalized contraction cell (which is not a wire), there is no reduction for $m$ (there could be reductions involving a coweakening but which would eventually erase $m$), and we do not reach an $(l,m)$-communication redex, this contradicts the hypotheses.
\item If it is the principal port of a prefix cell labeled $k$ ($k\neq l$), there is no $\{l,m\}$-reduction for $m$, and we do not reach an $(l,m)$-communication redex, this contradicts the hypotheses.
\item If it is the principal port of a generalized cocontraction cell of arity $0$, reductions involving $m$ will eventually erase it and we will never reach an $(l,m)$-communication redex, this contradicts the hypotheses.
\item\label{casempcontr} If it is the principal port of a generalized cocontraction cell of arity $n>0$, in order to reach an $(l,m)$-communication redex, the reductions involving $m$ must start destroying the generalized cocontraction.
In order to reach an $(l,m)$-communication redex 
we must escape the four cases above.
This means that we go to case~\ref{casempcontr} again 
until we reach the simplest \ref{casempcontr} case: $m$ facing $l$ (note that we cannot have infinitely many consecutive \ref{casempcontr} cases since the size of the net decreases).
\end{enumerate}
We conclude that, in order to reach an $(l,m)$-communication redex, we must go through a sequence of configurations~\ref{caselpcontr}$^*$-\ref{caselm} (see Figure~\ref{figproofdiving1}) or~\ref{caselpcontr}$^*$-\ref{caselauxcontr}-\ref{casempcontr}$^+$ (see Figure~\ref{figproofdiving2}).
For $l$ and $m$ to face each other after such a sequence, they must be connected to the same communication area $C$ in $\DINsone$ and the reduction sequence corresponds to the forwarding of $l$ and $m$ through $C$.
\qed

\begin{prop}\label{propbisimtwo}
If $\DINsone\NetTrans lm\DINstwo_0$ in $\NetLTS$ and $\SDone\SDDINRel\DINsone$ with $\SDone\in\SDLTSac$, then there are a solo diagram $\SDtwo$ and a simple net $\DINstwo$ such that $\SDone\SDTrans lm\SDtwo$ in $\SDLTSac$, $\SDtwo\SDDINRel\DINstwo$ and $\DINstwo_0\NetRedD\DINstwo$.
\end{prop}

\proof
Notations coincide with Figure~\ref{figproofpropbisimone}.
By definition of $\DINsone\NetTrans lm\DINstwo_0$, there exist a simple net $\DINsone_1$ containing an $(l,m)$-communication redex and a net $\DINone$ such that $\DINsone\Reltrans{\NetRedRel{\{l,m\}}}\DINsone_1+\DINone$ and $\DINstwo_0$ is obtained from $\DINsone_1$ by reducing this $(l,m)$-communication redex.
By Lemma~\ref{lemdiving}, $\SDone$ contains two dual multiedges $e_1$ and $e_2$ connected to the same node and labeled $l$ and $m$.
Let $\DINstwo_1$ be the simple net obtained by applying to $\DINsone_1$ the prefix reduction starting with the reduction from $\DINsone_1$ to $\DINstwo_0$ (see Lemma~\ref{lemprefix}),
we first show that $\SDred{\SDone}{e_1,e_2}\SDDINRel\DINstwo_1$. By Lemma~\ref{lemdiving} again, the only way the $(l,m)$-communication redex can have been generated in $\DINsone_1$ is by the forwarding of the codereliction $l$ and the dereliction $m$ through the communication area (of order $p+2$) they are both connected to. This gives a prefix redex and a communication area of order $p$ (Lemma~\ref{lemdercoder}). By firing this prefix redex, we generate pairs of wires connecting communication areas which exactly correspond to the identification edges of $\SDred{\SDone}{e_1,e_2}$.

Let $\SDtwo$ be the solo diagram (without identifications) obtained by firing the redex between $e_1$ and $e_2$ in $\SDone$ (this is possible since it is an acyclic redex). We iteratively apply Lemma~\ref{lemredident} starting from $\SDred{\SDone}{e_1,e_2}$ and $\DINstwo_1$ until we reach $\SDtwo$. This is possible since $\SDone\in\SDLTSac$. Let $\DINstwo$ be the simple net obtained this way and corresponding to $\SDtwo$, we have $\DINstwo_0\NetRedD\DINstwo$.
\qed

To get our bisimulation result, we define $\SDDINRelD$ as the composition of $\SDDINRel$ and $\NetEqD$: $\SDone\SDDINRelD\DINsone$ if there exists a simple net $\DINsone_0$ such that $\SDone\SDDINRel\DINsone_0$ and $\DINsone_0\NetEqD\DINsone$.

\begin{thm}[Bisimulation]
$\SDDINRelD$ is a bisimulation between $\SDLTSac$ and $\NetLTS$.
\end{thm}

\proof
If $\SDone\SDTrans lm\SDtwo$ in $\SDLTSac$, $\SDone\SDDINRel\DINsone_0$ and $\DINsone_0\NetEqD\DINsone$, by Proposition~\ref{propbisimone} there are simple nets $\DINstwo$ and $\DINstwo_0$ such that $\DINsone_0\NetTrans lm\DINstwo_0$ in $\NetLTS$, $\SDtwo\SDDINRel\DINstwo$ and $\DINstwo_0\NetRedD\DINstwo$. By Lemma~\ref{lemma:neteqd-bisimulation}, there exists a simple net $\DINstwo_1$ such that $\DINsone\NetTrans lm\DINstwo_1$ and $\DINstwo_0\NetEqD\DINstwo_1$. As a consequence $\DINstwo\NetEqD\DINstwo_1$ thus $\SDtwo\SDDINRelD\DINstwo_1$.

Conversely, if $\DINsone\NetTrans lm\DINstwo_1$ in $\NetLTS$, $\SDone\SDDINRel\DINsone_0$ and $\DINsone_0\NetEqD\DINsone$, by Lemma~\ref{lemma:neteqd-bisimulation} there exists a simple net $\DINstwo_0$ such that $\DINsone_0\NetTrans lm\DINstwo_0$ and $\DINstwo_1\NetEqD\DINstwo_0$.
By Proposition~\ref{propbisimtwo}, there are a solo diagram $\SDtwo$ and a simple net $\DINstwo$ such that $\SDone\SDTrans lm\SDtwo$ in $\SDLTSac$, $\SDtwo\SDDINRel\DINstwo$ and $\DINstwo_0\NetRedD\DINstwo$. As a consequence $\DINstwo\NetEqD\DINstwo_1$ thus $\SDtwo\SDDINRelD\DINstwo_1$.

\begin{equation*}
  \begin{array}{ccccc}
    \rnode{G}{\SDone} & \SDDINRel & \rnode{s0}{\DINsone_0} & \NetEqD & \rnode{s}{\DINsone} \\[4ex]
                      &           & \rnode{t0}{\DINstwo_0} & \NetEqD & \rnode{t1}{\DINstwo_1} \\[0.5ex]
& & \;\;\wr\,_{\mathrm{d}} \\[0.5ex]
    \rnode{H}{\SDtwo} & \SDDINRel & \rnode{t}{\DINstwo}
  \end{array}
\end{equation*}
\psset{nodesep=3pt}
\ncline{->}{G}{H}\tlput{$_{\LabPair{l}{m}}$}
\ncline{->}{s0}{t0}\tlput{$_{\LabPair{l}{m}}$}
\ncline{->}{s}{t1}\trput{$_{\LabPair{l}{m}}$}
\qed

The rest of the paper will be devoted to the definition of a sub-calculus of the solos calculus
whose translation into solo diagrams lives inside $\SDLTSac$.

\section{The acyclic solos calculus}\label{secacsolos}

By following the approach of the previous section and by analyzing the translation of the $\pi$-calculus into the solos calculus, we define a sub-system of the solos calculus called the \emph{acyclic solos calculus}. The key (informal) properties of this calculus are:
\begin{enumerate}[$\bullet$]
\item \emph{expressiveness}: it contains the image of the $\pi$-calculus through the translation into solos (Sections~\ref{sectyptrans} and~\ref{secactrad}).
\item \emph{stability}: it is well-defined with respect to the reduction of solos since it is stable under reduction (Propositions~\ref{propsubjred} and~\ref{proppresacred}).
\item \emph{acyclicity}: the solo diagram associated with a term of the acyclic solos calculus is an acyclic solo diagram (and thus the bisimulation with differential interaction nets holds!) (Section~\ref{secacsdiag}).
\end{enumerate}

The definition of the acyclic solos calculus is given in two steps: first by means of a typing system assigning Send/Receive polarities to occurrences of names, second by structural constraints on typed terms. Intuitively, we restrict terms to a forest-like structure with reduction occurring only between roots of the trees of the forest (in fact we will deal with more general structures than forests). This is an abstraction of the forest structure induced by the sequentiality of a $\pi$-term on its translation into solos.
The Send/Receive polarities on their side represent the input/output asymmetry of the $\pi$-calculus in the solos setting.

\subsection{Types}
\label{sectyping}

We consider a system with only two types $V$ and $W$. We use $U$ for denoting either $V$ or $W$.

A \emph{typed term} is a term with scopes decorated with types: $\SoloNu {x^U} P$.

A typing judgment is of the shape $\Gamma\vdash P$ where $\Gamma$ contains typing declarations associating types with names and $P$ is a typed term.
The typing rules are:
\begin{gather*}
  \AXC{}
  \UIC{$\Gamma\vdash\SoloEmpty$}
  \DP
\qquad\qquad\qquad
  \AXC{$\Gamma\vdash P$}
  \AXC{$\Gamma\vdash Q$}
  \BIC{$\Gamma\vdash\SoloPar P Q$}
  \DP
\qquad\qquad\qquad
  \AXC{$\Gamma,x:U\vdash P$}
  \UIC{$\Gamma\vdash\SoloNu {x^U} P$}
  \DP
\\[2ex]
  \AXC{}
  \UIC{$\Gamma,x:V,y:V,z:W,z':W\vdash\SoloInOut x {zz'y}$}
  \DP
\\[2ex]
  \AXC{}
  \UIC{$\Gamma,x:W,z:W,y:V,y':V\vdash\SoloInOut x {zyy'}$}
  \DP
\end{gather*}
where $\SoloInOut x{}$ is either $\SoloIn x{}$ or $\SoloOut x{}$.

An intuitive way of understanding the types $V$ and $W$ is given by these two mutually recursive definitions:
\begin{align*}
  V &::= WWV \\
  W &::= WVV
\end{align*}
(more generally, starting from other mutually recursive definitions of a finite set of types, one could derive other typing systems of the same kind which would also satisfy the subject reduction property for example).

The main purpose of this typing system is to be able to associate a \emph{communication protocol} with each object occurrence. Communication protocols are \ProtE{} (send) and \ProtR{} (receive). If we add communication protocols as a decoration on the definitions of $V$ and $W$:
\begin{align*}
  V &::= W^\ProtR W^\ProtE V^\ProtE \\
  W &::= W^\ProtR V^\ProtE V^\ProtR
\end{align*}
then any object occurrence of a typed term in context can be decorated with a communication protocol: in an input solo whose subject has type $U$, the decoration is given according to those of the components of the type, in an output solo it is given in a dual way: $\ProtE\mapsto\ProtR$ and $\ProtR\mapsto\ProtE$.
More formally, we can directly enrich typing derivations in such a way that they assign communication protocols to object occurrences of names in a typed term:
\begin{gather*}
  \AXC{}
  \UIC{$\Gamma\vdash\SoloEmpty$}
  \DP
\qquad\qquad\qquad
  \AXC{$\Gamma\vdash P$}
  \AXC{$\Gamma\vdash Q$}
  \BIC{$\Gamma\vdash\SoloPar P Q$}
  \DP
\qquad\qquad\qquad
  \AXC{$\Gamma,x:U\vdash P$}
  \UIC{$\Gamma\vdash\SoloNu {x^U} P$}
  \DP
\\[2ex]
  \AXC{}
  \UIC{$\Gamma,x:V,y:V,z:W,z':W\vdash\SoloIn x {z^\ProtR z'^\ProtE y^\ProtE}$}
  \DP
\qquad\qquad\qquad
  \AXC{}
  \UIC{$\Gamma,x:V,y:V,z:W,z':W\vdash\SoloOut x {z^\ProtE z'^\ProtR y^\ProtR}$}
  \DP
\\[2ex]
  \AXC{}
  \UIC{$\Gamma,x:W,z:W,y:V,y':V\vdash\SoloIn x {z^\ProtR y^\ProtE y'^\ProtR}$}
  \DP
\qquad\qquad\qquad
  \AXC{}
  \UIC{$\Gamma,x:W,z:W,y:V,y':V\vdash\SoloOut x {z^\ProtE y^\ProtR y'^\ProtE}$}
  \DP
\end{gather*}
Examples are given by the image of the translation of the $\pi$-calculus in Section~\ref{sectyptrans}.

The intuition behind \ProtE{} and \ProtR{} comes from the difference between the name-passing mechanism in the $\pi$-calculus and in the solos calculus. Name-passing by substitution (as in $\pi$) entails that a flow of information is given by interaction from an output prefix to an input prefix (receiving names are substituted with sending names). Name-passing by unification (as in fusion or solos) establishes a perfect symmetry between the agents trying to communicate. Communication protocols \ProtE{} and \ProtR{} are used to trace the flow of information coming from $\pi$ after translation into solos. Even if communication is symmetric (inside the solos calculus), it is reasonable for a term decorated with communication protocols to understand the unification of $x^\ProtE$ with $y^\ProtR$ as $y$ being substituted with $x$ (this will be a consequence of the constraints given in Definition~\ref{defacsolos}).

\subsubsection{Preservation by reduction}

We first check elementary properties of the typing system with respect to reduction.

\begin{lem}[Free names]
  If $\Gamma\vdash P$ then all the free names of $P$ appear in $\Gamma$.
\end{lem}

\begin{lem}[Weakening]\label{lemweak}
  If $x\notin\Gamma$ and $x\notin\FreeNames P$, $\Gamma\vdash P$ if and only if $\Gamma,x:U\vdash P$.
\end{lem}

\begin{lem}[Substitution]\label{lemsubst}
  If $\Gamma,x:U,y:U\vdash P$ then $\Gamma,x:U\vdash\Subst{P}{x}{y}$.
\end{lem}

\proof
  These three lemmas are proved by simple inductions on the typing derivations.
\qed

\begin{prop}[Subject reduction]\label{propsubjred}\hfill
  \begin{enumerate}[\em(1)]
  \item If $P\SoloCong Q$ then $\Gamma\vdash P\;\Leftrightarrow\;\Gamma\vdash Q$.
  \item If $P$ reduces to $Q$ and $\Gamma\vdash P$ then $\Gamma\vdash Q$.
  \end{enumerate}
\end{prop}

\proof\hfill
  \begin{enumerate}[(1)]
  \item We simply have to consider each case of the definition of $\SoloCong$ (given in Section~\ref{secsolos}). The only interesting one is $\SoloPar{(\SoloNu xP)}Q \SoloCong \SoloNu x{(\SoloPar PQ)}$ with $x\notin\FreeNames Q$. We assume $\Gamma\vdash\SoloPar{(\SoloNu {x^U}P)}Q$, this entails $\Gamma\vdash\SoloNu {x^U}P$ and $\Gamma\vdash Q$ and thus $\Gamma,x:U\vdash P$. By Lemma~\ref{lemweak}, we have $\Gamma,x:U\vdash Q$ thus $\Gamma,x:U\vdash \SoloPar PQ$ and finally $\Gamma\vdash\SoloNu {x^U}{(\SoloPar PQ)}$. Conversely, if $\Gamma\vdash\SoloNu {x^U}{(\SoloPar PQ)}$ then $\Gamma,x:U\vdash \SoloPar PQ$, $\Gamma,x:U\vdash P$, $\Gamma,x:U\vdash Q$ thus, by Lemma~\ref{lemweak} with $x\notin\FreeNames Q$, $\Gamma\vdash Q$ and:
    \begin{prooftree}
      \AXC{$\Gamma,x:U\vdash P$}
      \UIC{$\Gamma\vdash \SoloNu {x^U}P$}
      \AXC{$\Gamma\vdash Q$}
      \BIC{$\Gamma\vdash\SoloPar{(\SoloNu {x^U}P)}Q$}
    \end{prooftree}
  \item The only interesting case is $\SoloNu {\Many z}{(\SoloPar{\SoloPar{\SoloOut u{x_1x_2x_3}}{\SoloIn u{y_1y_2y_3}}}P)} \SoloRed P\sigma$ where $\Many z$ contains exactly the names such that $\sigma(w)\neq w$. From the hypothesis, we deduce $\Gamma,\Many z:\Many U\vdash \SoloOut u{x_1x_2x_3}$ and $\Gamma,\Many z:\Many U\vdash \SoloIn u{y_1y_2y_3}$ and $\Gamma,\Many z:\Many U\vdash P$. By Lemma~\ref{lemsubst}, we have $\Gamma\vdash P\sigma$.
\qed
  \end{enumerate}

\subsubsection{Typability of the translation}\label{sectyptrans}

We prove the typability of the translation of any $\pi$-term (see Section~\ref{sectranspisolos} for the definition of the translation).

If $P$ is a $\pi$-term with free names in the finite set $\mathcal{X}=\{x_1,\dots,x_k\}$, we define $\Gamma_{\mathcal{X}}$ to be the environment $x_1:W,\dots,x_k:W$ and $\Gamma_{\mathcal{X},v}$ to be the environment $\Gamma_{\mathcal{X}},v:V$. 

\begin{prop}[Typability]\label{proptyp}
  If $P$ is a $\pi$-term with free names in the finite set $\mathcal{X}$, $\Gamma_{\mathcal{X}}\vdash\ProcTrad{P}{}$.
\end{prop}

\proof
We first show that $\Gamma_{\mathcal{X},v}\vdash\ProcTrad{P}{v}$.

The process $\SoloCat v$ is typable:
\begin{prooftree}
\AXC{}
\UIC{$\Gamma_{\mathcal{X},v},z:W\vdash\SoloIn v {z^\ProtR z^\ProtE v^\ProtE}$}
\UIC{$\Gamma_{\mathcal{X},v}\vdash\SoloNu {z^W} {\SoloIn v {z^\ProtR z^\ProtE v^\ProtE}}$}
\end{prooftree}

The translations of prefixes are typable (see Figure~\ref{figtranspref}).

\begin{sidewaysfigure}
\centering
{\small
\begin{prooftree}
\AXC{}
\UIC{$\Gamma_{\mathcal{X},v},y:V,w:W\vdash\SoloOut v {u^\ProtE w^\ProtR y^\ProtR}$}
\AXC{$\Gamma_{\mathcal{X},v},y:V,w:W\vdash\SoloCat y$}
\BIC{$\Gamma_{\mathcal{X},v},y:V,w:W\vdash\SoloPar{\SoloOut v {u^\ProtE w^\ProtR y^\ProtR}}{\SoloCat y}$}
\AXC{}
\UIC{$\Gamma_{\mathcal{X},v},y:V,w:W,v':V,x:W\vdash\SoloIn w {x^\ProtR v^\ProtE v'^\ProtR}$}
\AXC{$\Gamma_{\mathcal{X},v},y:V,w:W,v':V,x:W\vdash\ProcTrad{P}{v'}$}
\BIC{$\Gamma_{\mathcal{X},v},y:V,w:W,v':V,x:W\vdash\SoloPar{\SoloIn w {x^\ProtR v^\ProtE v'^\ProtR}}{\ProcTrad{P}{v'}}$}
\UIC{$\Gamma_{\mathcal{X},v},y:V,w:W,x:W\vdash\SoloNu {v'^V}{(\SoloPar{\SoloIn w {x^\ProtR v^\ProtE v'^\ProtR}}{\ProcTrad{P}{v'}})}$}
\UIC{$\Gamma_{\mathcal{X},v},y:V,w:W\vdash\SoloNu {x^{W}}{\SoloNu {v'^V}{(\SoloPar{\SoloIn w {x^\ProtR v^\ProtE v'^\ProtR}}{\ProcTrad{P}{v'}})}}$}
\BIC{$\Gamma_{\mathcal{X},v},y:V,w:W\vdash\SoloPar{\SoloPar{\SoloOut v {u^\ProtE w^\ProtR y^\ProtR}}{\SoloCat y}}{\SoloNu {x^{W}}{\SoloNu {v'^V}{(\SoloPar{\SoloIn w {x^\ProtR v^\ProtE v'^\ProtR}}{\ProcTrad{P}{v'}})}}}$}
\UIC{$\Gamma_{\mathcal{X},v},w:W\vdash\SoloNu {y^V}{\SoloPar{\SoloPar{\SoloOut v {u^\ProtE w^\ProtR y^\ProtR}}{\SoloCat y}}{\SoloNu {x^{W}}{\SoloNu {v'^V}{(\SoloPar{\SoloIn w {x^\ProtR v^\ProtE v'^\ProtR}}{\ProcTrad{P}{v'}})}}}}$}
\UIC{$\Gamma_{\mathcal{X},v}\vdash\SoloNu {w^W}{\SoloNu {y^V}{\SoloPar{\SoloPar{\SoloOut v {u^\ProtE w^\ProtR y^\ProtR}}{\SoloCat y}}{\SoloNu {x^{W}}{\SoloNu {v'^V}{(\SoloPar{\SoloIn w {x^\ProtR v^\ProtE v'^\ProtR}}{\ProcTrad{P}{v'}})}}}}}$}
\end{prooftree}

\vspace{15ex}

\begin{prooftree}
\AXC{}
\UIC{$\Gamma_{\mathcal{X},v},y:V,w:W\vdash\SoloOut v {u^\ProtE w^\ProtR y^\ProtR}$}
\AXC{$\Gamma_{\mathcal{X},v},y:V,w:W\vdash\SoloCat y$}
\BIC{$\Gamma_{\mathcal{X},v},y:V,w:W\vdash\SoloPar{\SoloOut v {u^\ProtE w^\ProtR y^\ProtR}}{\SoloCat y}$}
\AXC{}
\UIC{$\Gamma_{\mathcal{X},v},y:V,w:W,v':V\vdash\SoloOut w {x^\ProtE v'^\ProtR v^\ProtE}$}
\AXC{$\Gamma_{\mathcal{X},v},y:V,w:W,v':V\vdash\ProcTrad{P}{v'}$}
\BIC{$\Gamma_{\mathcal{X},v},y:V,w:W,v':V\vdash\SoloPar{\SoloOut w {x^\ProtE v'^\ProtR v^\ProtE}}{\ProcTrad{P}{v'}}$}
\UIC{$\Gamma_{\mathcal{X},v},y:V,w:W\vdash\SoloNu {v'^V}{(\SoloPar{\SoloOut w {x^\ProtE v'^\ProtR v^\ProtE}}{\ProcTrad{P}{v'}})}$}
\BIC{$\Gamma_{\mathcal{X},v},y:V,w:W\vdash\SoloPar{\SoloPar{\SoloOut v {u^\ProtE w^\ProtR y^\ProtR}}{\SoloCat y}}{\SoloNu {v'^V}{(\SoloPar{\SoloOut w {x^\ProtE v'^\ProtR v^\ProtE}}{\ProcTrad{P}{v'}})}}$}
\UIC{$\Gamma_{\mathcal{X},v},w:W\vdash\SoloNu {y^V}{\SoloPar{\SoloPar{\SoloOut v {u^\ProtE w^\ProtR y^\ProtR}}{\SoloCat y}}{\SoloNu {v'^V}{(\SoloPar{\SoloOut w {x^\ProtE v'^\ProtR v^\ProtE}}{\ProcTrad{P}{v'}})}}}$}
\UIC{$\Gamma_{\mathcal{X},v}\vdash\SoloNu {w^W}{\SoloNu {y^V}{\SoloPar{\SoloPar{\SoloOut v {u^\ProtE w^\ProtR y^\ProtR}}{\SoloCat y}}{\SoloNu {v'^V}{(\SoloPar{\SoloOut w {x^\ProtE v'^\ProtR v^\ProtE}}{\ProcTrad{P}{v'}})}}}}$}
\end{prooftree}
}

\vspace{5ex}

\caption{Typing derivations for the translations of prefixes.}\label{figtranspref}
\end{sidewaysfigure}

The cases of $\ProcPar PQ$ and $\ProcNu xP$ are immediate.

Finally we get the typability of $\ProcTrad{P}{}$:
\begin{prooftree}
\AXC{$\Gamma_{\mathcal{X},v}\vdash\ProcTrad{P}{v}$}
\AXC{$\Gamma_{\mathcal{X},v}\vdash\SoloCat v$}
\BIC{$\Gamma_{\mathcal{X},v}\vdash\SoloPar{\ProcTrad{P}{v}}{\SoloCat v}$}
\UIC{$\Gamma_{\mathcal{X}}\vdash\SoloNu {v^V}{(\SoloPar{\ProcTrad{P}{v}}{\SoloCat v})}$}
\end{prooftree}
\qed

We can sum up the induced communication protocols:
\begin{align*}
  \SoloCat v &= \SoloNu {z^W} {\SoloIn v {z^\ProtR z^\ProtE v^\ProtE}} \\
  \ProcTrad{\ProcIn u{x}P}{v} &= \SoloNu {w^W}{\SoloNu {y^V}{(\SoloPar{\SoloOut v{u^\ProtE w^\ProtR y^\ProtR}}{\SoloPar{\SoloCat y}{\SoloNu{x^W}{\SoloNu{v'^V}{(\SoloPar{\SoloIn w{x^\ProtR v^\ProtE v'^\ProtR}}{\ProcTrad{P}{v'}})}}}})}} \\
  \ProcTrad{\ProcOut u{x}P}{v} &= \SoloNu {w^W}{\SoloNu {y^V}{(\SoloPar{\SoloOut v{u^\ProtE w^\ProtR y^\ProtR}}{\SoloPar{\SoloCat y}{\SoloNu{v'^V}{(\SoloPar{\SoloOut w{x^\ProtE v'^\ProtR v^\ProtE}}{\ProcTrad{P}{v'}})}}})}} \\
  \ProcTrad{\ProcPar{P}{Q}}{v} &= \SoloPar{\ProcTrad{P}{v}}{\ProcTrad{Q}{v}} \\
  \ProcTrad{\ProcNu x{P}}{v} &= \SoloNu {x^W}{\ProcTrad{P}{v}} \\
  \ProcTrad{P}{} &= \SoloNu {v^V}{(\SoloPar{\ProcTrad{P}{v}}{\SoloCat v})}
\end{align*}

\subsection{Acyclicity}\label{secsolosac}

Relying on the decoration with communication protocols induced by the typing system, we are now able to define our restriction of the solos calculus.

Given a typed term $P$, we first define some properties and relations between names and solos occurring in $P$.
If $s$ is a solo, we write $\Subject s$ for its subject.
\begin{equation*}
\begin{array}{rcl}
  x\in s          & \quad:=\quad & \textit{$x$ has an object occurrence in $s$} \\
  x^\ProtX\in s    & \quad:=\quad & \textit{$x$ has an object occurrence in $s$ with communication protocol $\ProtX$} \\
  s\SoloControl x & \quad:=\quad & x^\ProtR\in s \\
  s\SoloControl t & \quad:=\quad & s\SoloControl\Subject t \\
  s\SoloOrth t    & \quad:=\quad & \textit{$\Subject s=\Subject t$, one is output and the other is input} \\
  s\textit{ is a root} & \quad:=\quad & \textit{there is no $t$ such that $t\SoloControl s$}
\end{array}
\end{equation*}
We denote by $\SoloControlT$ the transitive closure of $\SoloControl$ and by $\SoloControlRT$ the reflexive transitive closure of $\SoloControl$.

\begin{defi}[Acyclic solos calculus]\label{defacsolos}
\emph{Acyclic solos terms} are typed terms which satisfy the following five properties:
\begin{center}
\fbox{
\begin{minipage}{8cm}
\begin{enumerate}
\renewcommand{\theenumi}{AC\arabic{enumi}}
\item\label{enumlinear} each name has at most one $\ProtR$-occurrence
\item\label{enumlock} $s\SoloOrth t$ implies that $s$ and $t$ are roots
\item\label{enumcontrol} $s\SoloControl x$ and $x\in t$ implies $s\SoloControlRT t$
\item\label{enumequal} $x^\ProtE\in s$ and $s\SoloControl x$ implies $s$ is an input
\item\label{enumbound} there is no free name with an $\ProtR$-occurrence
\end{enumerate}
\end{minipage}
}
\end{center}
The \emph{acyclic solos calculus} is given by restricting terms to acyclic solos terms. The structural congruence and the reduction relation are those induced by the solos calculus.
\end{defi}

We make a few remarks and state immediate consequences of this definition:
\begin{enumerate}[(1)]
\item Let us consider the following definition of communication protocols for occurrences of names in a $\pi$-term: if the occurrence is an object occurrence in an input prefix, the communication protocol is $\ProtR$, otherwise it is $\ProtE$. Assuming that the names appearing in a $\pi$-term are made as different as possible by $\alpha$-conversion, we can remark that each name has at most one $\ProtR$-occurrence, and no free name has an $\ProtR$-occurrence.
\item (\ref{enumlock}) entails that reduction only occurs between roots.
\item In the case where $\SoloControl$ induces a forest ordering, (\ref{enumcontrol}) means that all the $\ProtE$-occurrences of $x$ are bigger (with respect to this forest ordering) than its $\ProtR$-occurrence (if any).
\item (\ref{enumequal}) says that a name can have both an $\ProtE$-occurrence and an $\ProtR$-occurrence in a solo $s$ only if $s$ is an input. This is a place where we are breaking the symmetry of the solos calculus between output solos and input solos. This allows us to rule out processes like $\SoloPar{\SoloIn x{y^\ProtE y^\ProtR u^\ProtR}}{\SoloOut x{z^\ProtR z^\ProtE v^\ProtE}}$ that would lead to twice the identification of $y$ with $z$ which is almost the same as identifying $y$ with $y$ (and would break acyclicity of the associated solo diagram).
\item (\ref{enumbound}) ensures (with (\ref{enumlinear})) that identification edges are always contractible (no freeness problem).
\end{enumerate}

\subsubsection{Preservation by reduction}

We want to prove that reducts of an acyclic solos term under the reduction of the solos calculus are all acyclic solos terms (Proposition~\ref{proppresacred}).
We consider a reduction from $P$ to $Q$ by an interaction between the solos $s_0$ and $t_0$ of $P$. By definition of reduction (see Section~\ref{secsolos}), we have a substitution $\sigma$ such that each solo $t$ of $Q$ is of the shape $s\sigma$ for some solo $s$ of $P$. We say that $t$ is a \emph{residue} of $s$. The solos $s_0$ and $t_0$ of $P$ have no residue in $Q$ while all the other solos of $P$ have a unique residue in $Q$.
In this section, in order to simplify the notations, when $P$ reduces to $Q$ we will use the same name $s$ for a solo in $P$ and for its unique residue in $Q$ if it exists.

To make the difference between the relations $\SoloControl$ in different processes, we use the notation $\SoloControl_P$ for the $\SoloControl$ relation of the process $P$.

The following three lemmas hold assuming only conditions~(\ref{enumlinear}),~(\ref{enumlock}) and~(\ref{enumcontrol}) on $P$.

\begin{lem}\label{lemreceptsubst}
If $P$ is a typed term of the solos calculus satisfying (\ref{enumlinear}),~(\ref{enumlock}) and~(\ref{enumcontrol}) and which reduces to $Q$, 
then no $\ProtR$-occurrence of $Q$ has been substituted during reduction.
\end{lem}

\proof
Let $s_0$ and $t_0$ be the solos destroyed during the reduction from $P$ to $Q$, if a name $x$ involved in the reduction (\ie{} taking part in the unification, \ie{} occurring in $s_0$ or $t_0$) has its $\ProtR$-occurrence in a solo $t$ of $P$ different from $s_0$ and $t_0$ then $t\SoloControl_P x$ and, by~(\ref{enumcontrol}), either $t\SoloControlT_P s_0$ or $t\SoloControlT_P t_0$ thus one of them is not a root of $P$ contradicting~(\ref{enumlock}).
\qed

\begin{lem}\label{lemmodifsubj}
If $P$ is a typed term of the solos calculus satisfying (\ref{enumlinear}),~(\ref{enumlock}) and~(\ref{enumcontrol}) and which reduces to $Q$, 
if $s$ is a solo in $P$ whose subject has been modified during this reduction (\ie{} the subject of $s$ in $P$ is not the same as the subject of its residue in $Q$), we have both:
\begin{enumerate}[$\bullet$]
\item (the residue of) $s$ is a root in $Q$
\item $s$ is a root in $P$ or $s_0\SoloControl_P s$ or $t_0\SoloControl_P s$
\end{enumerate}
\end{lem}

\proof
Let $x$ be the subject of $s$ in $P$ and $y$ be its subject in $Q$, each of $x$ and $y$ appears in either $s_0$ or $t_0$. If $s$ is not a root in $Q$ then there is some $t$ such that $t\SoloControl_Q s$ thus $t\SoloControl_P y$ (by Lemma~\ref{lemreceptsubst}). If $y\in s_0$ (and symmetrically if $y\in t_0$) then $t\SoloControlRT_P s_0$ (by~(\ref{enumcontrol})) with $t\neq s_0$ since $t$ belongs to $Q$ thus $s_0$ is not a root in $P$ contradicting~(\ref{enumlock}). So that $s$ is a root in $Q$.

If $s$ is not a root in $P$ then there is some $t$ such that $t\SoloControl_P s$. Since the subject of $s$ is modified, it means that an $\ProtR$-occurrence in $t$ is also substituted thus $t\notin Q$ by Lemma~\ref{lemreceptsubst} and $t=s_0$ or $t=t_0$.
\qed

\begin{lem}\label{lemmodifrecept}
If $P$ is a typed term of the solos calculus satisfying (\ref{enumlinear}),~(\ref{enumlock}) and~(\ref{enumcontrol}) and which reduces to $Q$, 
if the reduction introduces an object occurrence of the name $x$ in $Q$ then $x$ has no $\ProtR$-occurrence in $Q$.
\end{lem}

\proof
For the reduction to put some $x$ in some $t$, $x$ must occur in $s_0$ or $t_0$. Assume that $x$ occurs in $s_0$ and that $x$ has an $\ProtR$-occurrence in $s$ in $Q$ (so that $s\SoloControl_Q x$). By Lemma~\ref{lemreceptsubst}, $s\SoloControl_P x$ and by~(\ref{enumcontrol}), we have $s\SoloControlRT_P s_0$ but, since $s\neq s_0$ (because $s_0\notin Q$), this would entail that $s_0$ is not a root in $P$ contradicting~(\ref{enumlock}).
\qed

We now turn to the preservation result.

\begin{prop}[Acyclicity preservation]\label{proppresacred}
If $P$ is an acyclic solos term and $P$ reduces to $Q$ then $Q$ is an acyclic solos term.
\end{prop}

\proof
We first prove the preservation of conditions~(\ref{enumlinear}),~(\ref{enumlock}) and~(\ref{enumcontrol}), independently of conditions~(\ref{enumequal}) and~(\ref{enumbound}).

Preservation of condition~(\ref{enumlinear}) is given by Lemma~\ref{lemreceptsubst}.

We move to condition~(\ref{enumlock}). If $s\SoloOrth_Q s'$, note that if one of them is not a root then none of them is since $t\SoloControl s\Rightarrow t\SoloControl\Subject s\Rightarrow t\SoloControl\Subject{s'}\Rightarrow t\SoloControl s'$. Assume that $t\SoloControl_Q s$ and $t\SoloControl_Q s'$, if the subject of neither $s$ nor $s'$ has been modified during reduction, then $t\SoloControl_P s$ and $t\SoloControl_P s'$ (since the $\ProtR$-occurrence in $t$ of the subject of $s$ and $s'$ has not been modified, by Lemma~\ref{lemreceptsubst}) and $P$ contradicts condition~(\ref{enumlock}). If one of the subjects of $s$ or $s'$ has been modified then it is a root in $Q$ by Lemma~\ref{lemmodifsubj}.

Concerning condition~(\ref{enumcontrol}), we assume $s\SoloControl_Q x$ and $x\in_Q t$. By Lemma~\ref{lemreceptsubst}, $s\SoloControl_P x$ and by Lemma~\ref{lemmodifrecept}, $x\in_P t$.
So that, by~(\ref{enumcontrol}), $s\SoloControlRT_P t$. If $s{\not\!\SoloControlRT_Q} t$, then the path from $s$ to $t$ for the relation $\SoloControl_P$ has been broken during reduction, this entails that a solo $t'\neq s$ in this path has got his subject modified. By Lemma~\ref{lemmodifsubj}, $s_0\SoloControl_P t'$ or $t_0\SoloControl_P t'$ but this is impossible because, since $s_0$ and $t_0$ are roots in $P$ and are not in $Q$, $s\neq s_0$ and $s\neq t_0$.

We now assume that~(\ref{enumlinear}),~(\ref{enumlock}) and~(\ref{enumcontrol}) hold. Condition~(\ref{enumequal}) is preserved: if, in $Q$, an output solo $s$ contains both an $\ProtR$-occurrence and an $\ProtE$-occurrence of a name $x$ then, by Lemmas~\ref{lemreceptsubst} and~\ref{lemmodifrecept}, it is also the case in $P$. Condition~(\ref{enumbound}) is preserved by Lemma~\ref{lemreceptsubst}.
\qed

All this shows that our acyclic solos calculus is well behaved with respect to the reduction of the solos calculus.

\subsubsection{Acyclicity of the translation}\label{secactrad}

In order to prove the expressiveness of the acyclic solos calculus, we check that it contains the image of the translation of the $\pi$-calculus.

The $\SoloControl$ relation is the following for translated $\pi$-terms:
\begin{align*}
  \SoloCat v &= \SoloNu {z^W} {\SoloIn v {z^\ProtR z^\ProtE v^\ProtE}} \\
  \ProcTrad{\ProcEmpty}{v} &= \SoloEmpty \\
  \ProcTrad{\ProcIn u{x}P}{v} &= \SoloNu {w^W}{\SoloNu {y^V}{(\SoloPar{\rnode{v1}{\SoloOut v{u^\ProtE w^\ProtR y^\ProtR}}}{\SoloPar{\rnode{cat1}{\SoloCat y}}{\SoloNu{x^W}{\SoloNu{v'^V}{(\SoloPar{\rnode{w1}{\SoloIn w{x^\ProtR v^\ProtE v'^\ProtR}}}{\rnode{p1}{\ProcTrad{P}{v'}}})}}}})}} \\
  \ProcTrad{\ProcOut u{x}P}{v} &= \SoloNu {w^W}{\SoloNu {y^V}{(\SoloPar{\rnode{v2}{\SoloOut v{u^\ProtE w^\ProtR y^\ProtR}}}{\SoloPar{\rnode{cat2}{\SoloCat y}}{\SoloNu{v'^V}{(\SoloPar{\rnode{w2}{\SoloOut w{x^\ProtE v'^\ProtR v^\ProtE}}}{\rnode{p2}{\ProcTrad{P}{v'}}})}}})}} \\
  \ProcTrad{\ProcPar{P}{Q}}{v} &= \SoloPar{\ProcTrad{P}{v}}{\ProcTrad{Q}{v}} \\
  \ProcTrad{\ProcNu x{P}}{v} &= \SoloNu {x^W}{\ProcTrad{P}{v}} \\
  \ProcTrad{P}{} &= \SoloNu{v^V}{(\SoloPar{\ProcTrad{P}{v}}{\SoloCat v})}
\end{align*}
where an arrow $\rnode{s}{s}\qquad\rnode{t}{t}$ represents the relation $s\SoloControl t$.
\ncarc[nodesep=1pt,arcangle=-40,arrowscale=2]{->}{t}{s}
\ncarc[nodesep=1pt,arcangle=-50,arrowscale=2]{->}{cat1}{v1}
\ncarc[nodesep=1pt,arcangle=-50,arrowscale=2]{->}{w1}{v1}
\ncarc[nodesep=1pt,arcangle=-50,arrowscale=2]{->}{p1}{w1}
\ncarc[nodesep=1pt,arcangle=50,arrowscale=2]{->}{cat2}{v2}
\ncarc[nodesep=1pt,arcangle=50,arrowscale=2]{->}{w2}{v2}
\ncarc[nodesep=1pt,arcangle=50,arrowscale=2]{->}{p2}{w2}

\begin{thm}[Acyclicity for the $\pi$-calculus]
  If $P$ is a $\pi$-term, $\ProcTrad{P}{}$ is an acyclic solos term.
\end{thm}

\proof
By Proposition~\ref{proptyp}, $\ProcTrad{P}{}$ is typable.

Let us mention a few additional facts (which can be easily checked by induction on the definition of $\ProcTrad{P}{v}$):
\begin{enumerate}[(i)]
\item\label{ffnp} The free names of $\ProcTrad{P}{v}$ are $v$ and the free names of $P$.
\item\label{ffnr} The free names of $P$ have no $\ProtR$-occurrence in $\ProcTrad{P}{v}$.
\item\label{fsr} The only subject with no $\ProtR$-occurrence (thus the only subject of a root) in $\ProcTrad{P}{v}$ is $v$.
\item\label{ffor} $\SoloControl$ is a forest on $\ProcTrad{P}{v}$ (which is one of the reasons for the name ``acyclic solos'').
\end{enumerate}
From these points, conditions~(\ref{enumlinear}),~(\ref{enumlock}),~(\ref{enumcontrol}),~(\ref{enumequal}) and~(\ref{enumbound}) are easily verified:
\begin{enumerate}[$\bullet$]
\item We first check the five conditions for $\ProcTrad{P}{v}$:
\begin{enumerate}[(AC1)]
\item By induction on $\ProcTrad{P}{v}$ by using facts~\ref{ffnp}, \ref{ffnr} and~\ref{fsr} which entail that no free name in $\ProcTrad{P}{v}$ has an $\ProtR$-occurrence in $\ProcTrad{P}{v}$.
\item An easy induction with fact~\ref{ffnp} shows that $s\SoloOrth t$ entails $\Subject s=\Subject t=v$ and we conclude with fact~\ref{fsr}.
\item By induction, using facts~\ref{fsr} and~\ref{ffor}.
\item Immediate induction.
\item A consequence of facts~\ref{ffnr}, \ref{ffnp} and~\ref{fsr}.
\end{enumerate}
\item It is then easy to check the case of $\ProcTrad{P}{}$.
\qed
\end{enumerate}

\subsection{Back to acyclic solo diagrams}\label{secacsdiag}

The last property we want to show is that the solo diagram associated with a term of the acyclic solos calculus (Definition~\ref{defacsolos}) is an acyclic solo diagram (Definition~\ref{defacdiag}).

Let $\SDone$ be such a solo diagram associated with the term $P$ of the acyclic solos calculus, we have to show that any pair of dual edges in $\SDone$ is an acyclic redex (Lemma~\ref{lemacredex}). This will be enough to show that $\SDone$ belongs to $\SDLTSac$ (Theorem~\ref{thmac}).

\begin{lem}\label{lemacredex}
Let $\SDone$ be the solo diagram associated with an acyclic solos term $P$ and let $e_1$ and $e_2$ be two dual multiedges of $\SDone$, this pair of multiedges defines an acyclic redex (Definition~\ref{defacredex}).
\end{lem}

We want to show that the graph $\NdSDred{\SDone}{e_1,e_2}$ (sub-graph of $\SDred{\SDone}{e_1,e_2}$ with identification edges only, as introduced in Section~\ref{secrestrlts}) is acyclic and does not generate any freeness problem in the contraction of identification edges.

By construction, edges in $\NdSDred{\SDone}{e_1,e_2}$ are connecting two nodes which correspond to occurrences of names in the term $P$ which are going to be unified. By definition of the acyclic solos calculus, one of these occurrences is an $\ProtE$-occurrence and the other one is an $\ProtR$-occurrence. We consider the directed graph $\SDone'$  obtained by orienting the edges of $\NdSDred{\SDone}{e_1,e_2}$ towards $\ProtR$-occurrences.

We first prove the following lemma about directed and non-directed graphs.

\begin{lem}\label{lemgraph}
Let $g$ be a finite directed graph and $g_s$ be the underlying non-directed graph. If $g_s$ is connected, if $g$ has a root (that is a node without incoming edge) and if each node of $g$ has at most one incoming edge then $g_s$ is acyclic.
\end{lem}

\proof
The function which maps each edge of $g$ to its target is an injective function (by hypotheses) and its image is the set of nodes of $g$ which are not roots. As a consequence, the number of edges of $g$ (and thus of $g_s$) is less than or equal to the number of vertices of $g$ (and thus of $g_s$) minus one.
Since $g_s$ is connected, we conclude that it is acyclic.
\qed

We can now prove Lemma~\ref{lemacredex}:

\proof
We prove that the connected components of $\NdSDred{\SDone}{e_1,e_2}$ equipped with the orientation of $\SDone'$ satisfy the hypotheses of Lemma~\ref{lemgraph} and thus that $\NdSDred{\SDone}{e_1,e_2}$ is acyclic. Let us assume $e_1$ is the input multiedge in the redex between $e_1$ and $e_2$ in $\SDone$ (since solos in a solos term are in one-to-one correspondence with multiedges in the corresponding solo diagram, we use the notations introduced in the beginning of Section~\ref{secsolosac} with multiedges as well as with solos), we have the following properties:
\begin{enumerate}[(i)]
\item\label{caseR1} if $x^\ProtR\in e_1$ then all the other occurrences of $x$ in $e_1$ and $e_2$ are $\ProtE$-occurrences in $e_1$: by~(\ref{enumlinear}), $x$ cannot have another $\ProtR$-occurrence, moreover, if $x^\ProtE\in e_2$ then, by~(\ref{enumcontrol}), $e_1\SoloControlT e_2$ thus $e_2$ is not a root of $P$, contradicting~(\ref{enumlock}).
\item\label{caseR2} if $x^\ProtR\in e_2$ then $x$ has no other occurrence neither in $e_1$ nor in $e_2$: as for~\ref{caseR1}, $x$ cannot have any other $\ProtR$-occurrence nor any $\ProtE$-occurrence in $e_1$ and finally, by~(\ref{enumequal}), it cannot have an $\ProtE$-occurrence in $e_2$.
\item\label{caseE2} if $x^\ProtE\in e_2$ then $x$ has no $\ProtR$-occurrence neither in $e_1$ nor in $e_2$: otherwise we apply~\ref{caseR1} or~\ref{caseR2}.
\end{enumerate}
We consider a connected component $\SDone_0$ of $\NdSDred{\SDone}{e_1,e_2}$. It is the underlying non-directed graph of the appropriate sub-graph $\SDone'_0$ of $\SDone'$. $\SDone'_0$ is finite,  $\SDone_0$ is connected and, according to condition~(\ref{enumlinear}), each node has at most one incoming edge in $\SDone'_0$. We just have to show that $\SDone'_0$ has a root, \ie{} that there is a name in $P$ which has only $\ProtE$-occurrences in $e_1$ and $e_2$
for each connected component $\SDone_0$.
We first show that any $\ProtR$-occurrence in $e_2$ is connected to an $\ProtE$-occurrence of a name without $\ProtR$-occurrence: we start from this $\ProtR$-occurrence in $e_2$, the corresponding occurrence in $e_1$ is an $\ProtE$-occurrence of a name $x$, if $x$ has no $\ProtR$-occurrence we are done, and if $x$ has an $\ProtR$-occurrence then it is also in $e_1$ (by~\ref{caseR2}) and the corresponding occurrence in $e_2$ is an $\ProtE$-occurrence (thus without $\ProtR$-occurrence by~\ref{caseE2}). Since any connected component contains either an $\ProtR$-occurrence in $e_2$ or an $\ProtE$-occurrence in $e_2$, we are done (with~\ref{caseE2} again).

Finally by~(\ref{enumbound}), $\SDone'$ has the property that only roots can be free nodes (according to the free/bound labeling of nodes in $\SDone$). This property is preserved by contraction of identification edges thanks to~(\ref{enumlinear}). Moreover this property entails that at most one extremity of an identification edge can be free, and thus no freeness problem arises during contraction of identification edges.
\qed

\begin{thm}[Acyclic diagrams from acyclic terms]\label{thmac}
The solo diagram associated with an acyclic solos term (Definition~\ref{defacsolos}) is an acyclic solo diagram (Definition~\ref{defacdiag}).
\end{thm}

\proof
We prove that the range of the translation of the acyclic solos calculus into solo diagrams satisfies the conditions of Definition~\ref{defltssdac}.
By Lemma~\ref{lemacredex}, we know such a translation contains only acyclic redexes. Now if $\SDone$ is the translation of the acyclic solos term $P$ and reduces to $\SDtwo$, there exists a term $Q$ which translates into $\SDtwo$, and such that $P$ reduces to $Q$ thus $Q$ is an acyclic solos term (Proposition~\ref{proppresacred}).
\qed

\section*{Conclusion}

In the spirit of a Curry-Howard correspondence between a logical device and a concurrent one, we have shown how to stress a strong connection between differential interaction nets and the solos calculus. Such links allow one to share methods from the two worlds, with the possibility of giving concurrent interpretations to the cut-elimination procedure, logical foundations to process calculi, of applying tools from concurrency theory such as behavioural equivalences to formal proofs, or conversely denotational semantics from linear logic to concurrent languages, ...

Let us discuss a few constraints or restrictions we have to face here.

\paragraph{Finitary calculi.}
We have only considered finitary calculi: without replications nor recursive definitions. Nevertheless, \emph{exponential boxes} are a natural device from linear logic proof-nets (compatible with differential interaction nets) for representing replicable processes. We have shown in~\cite{pidinic} how to represent with these boxes a restricted (but expressive) form of replication of the $\pi$-calculus. This restriction is related with difficulties for controling reduction on the auxiliary ports of exponential boxes. By exploiting this kind of restriction in the language of solos, it should be possible to represent a restricted form of replication through our translation.

\paragraph{Logical correctness.}
Proof-nets of linear logic provide us with a graphical syntax for representing proofs of the sequent calculus~\cite{pn}. However this requires to impose a \emph{correctness criterion} to proof-nets in order to characterize exactly those which could be sequentialized into a sequent calculus proof. There is such a correctness criterion for differential interaction nets~\cite{din} (relating them with the sequent calculus of differential linear logic). However the differential interaction nets obtained with our translation do not satisfy the correctness criterion in general (this comes in particular from the unconstrained use of communication areas). Some logical tools do not care about logical correctness and can be directly applied in our setting (for example, the relational model is a denotational model of differential interaction nets even if they do not satisfy correctness). On the contrary, many important results in logic only hold in the logically correct case. This is why it would be important to understand how our translation interacts with correctness, or how expressive would be a concurrent calculus whose translation into differential interaction nets satisfies the correctness criterion. In particular, what kind of communications could be represented with communication areas in a logically correct setting?

\paragraph{Acyclicity.}
In trying to build a strong bridge between differential interaction nets and solo diagrams, 
we have been able to define a suitable restriction of the solos calculus that can be used as an intermediary step between the $\pi$-calculus and differential interaction nets. This led to a new proof of the bisimulation result presented in~\cite{pidinic}.

The technical choices in the design of our acyclic solos calculus were completely determined by the just mentioned goal. Nevertheless, it would be interesting to study acyclic solos for themselves. Their computational behaviour are in many ways similar to what happens in the $\pi$-calculus (``substitution-like'' name passing in an ``unification style'' setting for example). It would be interesting to see if they contain specific communication primitives or if somehow the behaviour of an acyclic solos term always mimics the behaviour of a $\pi$-term.

Another approach would be to extend our translation relation to take cycles (as in Figure~\ref{fig:redcomp}) into account. This requires us to understand the precise impact of these cycles on the behaviour of differential interaction nets. In this way, it seems possible to extend the bisimulation result to the whole solos calculus.

\bigskip

Following the link provided by the present work, the above mentioned points should now be addressed to strengthen the idea of an underlying Curry-Howard correspondence.
However some possibilities are already open such as relational semantics for the $\pi$-calculus, geometry of interaction for processes,~...

\section*{Acknowledgement}

We would like to thank Cosimo Laneve for the helpful discussions we had about solos.
We also thank the anonymous referees for their useful comments.

\bibliographystyle{alpha}
\bibliography{solos}

\end{document}